\documentclass{article}
\usepackage{graphicx}
\usepackage{amsmath}
\usepackage{amssymb}
\usepackage{amsfonts}
\usepackage{color}
\usepackage{subfigure}
\usepackage{cases}
\usepackage{setspace}
\usepackage{fullpage}

\newtheorem{definition}{Definition}

\newtheorem{proposition}{Proposition}

\def\theequation{{\arabic{section}}.{\arabic{equation}}}
\newcommand{\nc}{\newcommand}
\nc{\R}{\Bbb{R}}
\nc{\Z}{\Bbb{Z}}
\nc{\Pp}{\Bbb{P}}
\nc{\Ap}{\Bbb{A}}
\nc{\Wp}{\Bbb{W}}
 \nc{\brho}{\boldsymbol \rho}
\nc{\va}{\vec{\boldsymbol \mu}}
\nc{\ve}{\vec{\boldsymbol \epsilon}}
\nc{\bk}{{\bf k}}
\nc{\vrho}{\vec{\brho}} 
\nc{\vr}{\vec{\bf r}}
\nc{\bx}{{\bf x}}
\nc{\vx}{\vec{\bf x}}
\nc{\om}{\omega}
\nc{\brhoi}{\brho^{\cal I}}
\nc{\bzetai}{\bzeta^{\cal I}}
\nc{\vzeta}{\vec{\bzeta}}
\nc{\vzetai}{\vec{\bzeta}^{\cal I}}
\nc{\cI}{{\cal I}}
\nc{\cJ}{{\cal J}}
\nc{\cF}{{\cal F}}
\nc{\cW}{{\cal W}}
\nc{\cA}{{\cal A}}
\nc{\cL}{{\cal L}}
\nc{\cS}{{\cal S}}
\nc{\cC}{{\cal C}}
\nc{\cN}{{\cal N}}
\nc{\vrhoi}{\vec{\brho}^{\,\cal I}}
\nc{\xii}{\xi^{\cI}}
\nc{\etai}{\eta^{\cI}}
\nc{\la}{\lambda}
\nc{\de}{\delta}
\nc{\ep}{\varepsilon}
\nc{\vu}{\vec{\bf u}}
\nc{\bu}{{\bf u}}
\nc{\vui}{\vec{\bf u}^{\cI}}
\nc{\bui}{{\bf u}^{\cI}}
\nc{\bt}{{\bf t}}
\nc{\vt}{\vec{\bt}}
\nc{\bn}{{\bf n}}
\nc{\vn}{\vec{\bn}}
\nc{\bm}{{\bf m}}
\nc{\vm}{\vec{\bm}}
\nc{\vrp}{\vec{{\bf r}'}}
\nc{\vrc}{\vec{{\bf r}^c_p}}
\nc{\ts}{\tilde s}
\nc{\os}{\overline s}
\nc{\tom}{\tilde \om}
\nc{\tO}{\tilde \Omega}
\nc{\tS}{\tilde S}
\nc{\oS}{\overline S}
\nc{\vrhos}{\vrho_{\star}}
\nc{\vrhosi}{\vrho_{\star}^{\cI}}
\nc{\brhos}{\brho_{\star}}
\nc{\brhosi}{\brhos^{\cI}}
\nc{\vms}{\vm_{\star}}
\nc{\vmi}{\vm_{_{\cI}}}
\nc{\vM}{\vec{\bf M}}
\nc{\vMi}{\vM_{_{\cI}}}
\nc{\Ppi}{\Pp_{\cI}}
\nc{\vxi}{\vec{\bxi}}
\nc{\bK}{{\bf K}}
\nc{\bmi}{\bm_{_{\cI}}}
\nc{\Pppi}{\Ppi^p}
\nc{\be}{{\bf e}}
\nc{\bep}{{\bf e}^p}
\renewcommand{\hat}{\widehat}

\begin{document}
\title{Synthetic Aperture Radar Imaging and Motion Estimation \\
  via Robust Principal Component Analysis} \author{Liliana
  Borcea\footnotemark[2], Thomas Callaghan\footnotemark[2], and George
  Papanicolaou\footnotemark[3]
}\renewcommand{\thefootnote}{\fnsymbol{footnote}}
\footnotetext[2]{Computational and Applied Mathematics, Rice
  University, MS 134, Houston, TX 77005-1892. (borcea@caam.rice.edu
  and tscallaghan@rice.edu)} \footnotetext[3]{Department of
  Mathematics, Stanford University, Stanford, CA 94305.
  (papanicolalou@stanford.edu)} \date{}
  \maketitle 


\begin{abstract}
  We consider the problem of synthetic aperture radar (SAR) imaging
  and motion estimation of complex scenes. By complex we mean scenes
  with multiple targets, stationary and in motion. We use the usual
  setup with one moving antenna emitting and receiving signals. We
  address two challenges: (1) the detection of moving targets in the
  complex scene and (2) the separation of the echoes from the
  stationary targets and those from the moving targets.  Such
  separation allows high resolution imaging of the stationary scene
  and motion estimation with the echoes from the moving targets alone.
  We show that the robust principal component analysis (PCA) method
  which decomposes a matrix in two parts, one low rank and one sparse,
  can be used for motion detection and data separation. The matrix
  that is decomposed is the pulse and range compressed SAR data
  indexed by two discrete time variables: the slow time, which
  parametrizes the location of the antenna, and the fast time, which
  parametrizes the echoes received between successive emissions from
  the antenna. We present an analysis of the rank of the data matrix
  to motivate the use of the robust PCA method.  We also show with
  numerical simulations that successful data separation with robust
  PCA requires proper data windowing.  Results of motion estimation
  and imaging with the separated data are presented, as well.
\end{abstract}

\section{Introduction}
\label{sect:intro}
\setcounter{equation}{0}

\begin{figure}[t]
\vspace{-0.4in}
  \begin{center}
    \input{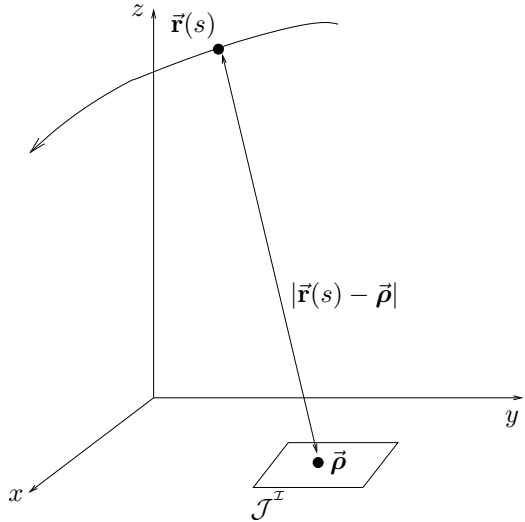}
  \end{center}
  \caption{Setup for synthetic aperture imaging.}
  \label{fig:setup}
\end{figure}

In synthetic aperture radar (SAR), the basic problem
\cite{cheney2001mathematical,Jakowatz, Curlander} is to image the
reflectivity supported in a set $\cJ^{^\cI}$ on the ground surface
using measurements obtained with an antenna system mounted on a
platform flying above it, as illustrated in Figure \ref{fig:setup}.
The antenna emits periodically a probing signal $f(t)$ and records
the echos $D(s,t)$, indexed by the slow time $s$ of the SAR platform
displacement and the fast time $t$. The slow time parametrizes the
location $\vr(s)$ of the platform at the instant it emits the signal,
and the fast time $t$ parametrizes the echoes received between two
consecutive illuminations $\left( 0 < t< \Delta s\right)$.
The echoes $D(s,t)$ are approximately, and up to a multiplicative
factor, a superposition of the emitted signals $f(t)$ time-delayed by
the round-trip travel-time between the platform $\vr(s)$ and the
locations $\vrho$ of scatterers on the ground
\begin{equation}
\tau\left(s,\vrho\right)=2\left|\vr(s)-\vrho\right|/c.
\end{equation}
Here $c$ is the wave speed, assumed constant and equal to the speed of
light.

A SAR image is formed by superposing over a platform trajectory of
length (aperture) $a$ the data $D(s,t)$ convolved with the time
reversed emitted signal (matched filtered), and then backpropagated to points in the
imaging domain $\cJ^{^\cI}$ using travel times. With high bandwidth
probing signals and large flight apertures, SAR is capable of
generating images with roughly ten centimeter resolution at ranges of
ten kilometers away from the platform.  Such resolution cannot be
achieved for complex scenes because moving targets appear blurred and
displaced in the images. Image formation should therefore be done in
conjunction with target motion estimation. In fact, in applications
such as persistent surveillance SAR, tracking and imaging the moving
targets is one of the primary objectives.

Because targets may have complicated motion over lengthy data
acquisition trajectories, the targets are tracked over successive small
sub-apertures. Each sub-aperture corresponds to a short time interval
over which the target is in approximate uniform translational
motion. The problem is to estimate this motion for each time interval
in order to bring the small aperture images of the moving targets into
focus. Then, the images are superposed to form high resolution images
over larger apertures.

The existing algorithms for motion estimation fall roughly into two
categories: The first is for the usual SAR setup with a single
moving antenna, and the target motion is estimated from the phase
modulations of the return echoes
\cite{sar,ding2000analysis,ding2002time,barbarossa1992detection,
  sparr-time, zhuMTI, fienup, jao, kirscht, perry,ender1993}.  These
algorithms assume that all the targets are in the same motion, and are
sensitive to the presence of strong stationary targets.  The second
class of methods uses more complex antenna systems \cite{friedlander,
  wang2004,wang2006}, with multiple receiver and/or transmitter
antennas. They form a collection of images with the echoes measured by
each receiver-transmitter pair, and then they use the phase variation
of the images with respect to the receiver/transmitter offsets, pixel
by pixel, to extract the target velocity.

In this paper we consider imaging and motion estimation of complex
scenes with the usual SAR setup, using a single antenna.  We address
two challenges: (1) the detection of moving targets in the scene and
(2) the separation of data in subsets of echoes from the stationary
scene and echoes from the moving targets.  The stationary scene can be
imaged by itself after such separation, and the motion estimation can
be carried out on the echoes from the moving targets alone.  We propose
and analyze a detection and data separation approach based on the
robust principle component analysis (robust PCA) method
\cite{candesRPCA}.  Robust PCA is designed to decompose a matrix into
a low rank one plus a sparse one.  The main contribution of this
paper is to show with analysis and numerical simulations that by
appropriately pre-processing and windowing the SAR data we can
decompose it into a low rank part, corresponding to the stationary
scene, and a sparse part, corresponding to the moving targets. Our
theoretical and numerical study describes the rank of the
pre-processed SAR data as a function of the velocity, location, and
density of the scatterers. It specifies in particular how slowly a target
can move and still be distinguishable from the stationary scene. It
also addresses the question of proper windowing of the data for the
separation with robust PCA to work.

The paper is organized as follows: We begin in section
\ref{sect:BasicSAR} with a brief description of basic SAR data
processing and image formation. There are two processing steps that
are key to the data decomposition: pulse compression and range
compression.  We illustrate with numerical simulations in section
\ref{sect:decomp} that robust PCA can be used for motion detection and
data separation if it is complemented with proper data windowing. The
analysis is in section \ref{sect:rank}.  Additional numerical results
on data separation, motion estimation and imaging are in section
\ref{sect:numerics}. We end with a summary in section \ref{sect:conc}.

\section{Basic SAR data processing and image formation}
\label{sect:BasicSAR}

The antenna emits signals that consist of a base-band waveform
$f_B(t)$ modulated by a carrier frequency $\nu_o = \om_o/(2 \pi)$,
\begin{equation}
f(t) =  \cos(\om_o t) f_B(t).
\label{eq:1.1}
\end{equation}
Its Fourier transform is
\begin{eqnarray}
  \hat f(\om) = \int d t \, f(t) e^{i \om t} 
  = \frac{1}{2} \left[\hat f_B(\om + \om_o) + 
    \hat f_B(\om - \om_o)\right],
\end{eqnarray}
with $\hat f_B(\om)$ supported in the interval $[-\pi B,\pi B]$, where
$B$ is the bandwidth. The support of $\hat f(\om)$ is the same
interval with center shifted at $\om_o$, and its mirror image in the
negative frequencies.

Imaging relies on accurate estimation of travel times. Recall that the
echoes are approximately, up to some amplitude factors, superpositions
of the emitted signals delayed by the round trip travel times between
the antenna and the targets in the scene. Suppose that
\[
f_B(t) = \varphi(Bt),
\]
with $\varphi$ a function of a dimensionless argument and compact
support in the unit interval $[0,1]$. Then $f_B(t)$ has the form of a
pulse, and the travel times can be estimated with precision $1/B$, the
pulse support.  But such pulses are almost never used in SAR because
of power limitations at the antenna \cite{Jakowatz}. The SAR echoes
should be above the antenna's noise level, so the emitted signals should
carry large power.  However, the instantaneous power at the antenna is
limited, while large net power can be delivered by longer signals,
such as chirps.  The problem is that the received scattered energy is
spread out over the long time support of such signals, making it
impossible to resolve the time of arrival of different echoes. This is
overcome by compressing the long echoes, as if they were created by an
incident pulse. This \emph{pulse compression} amounts to convolving
the data with the complex conjugate of the time reversed emitted
signal $f(-t)$
\begin{equation}
D_p(s,t) = \int dt' \, D(s,t') \overline{f \left(t' -t \right)}.
\end{equation}
The result is as if the antenna emitted the signal 
\[
f_p(t) = \int dt' \, f(t') \overline{f \left(t' -t \right)},
\]
which turns out to be a pulse of support $1/B$, as explained in detail
in \cite{Jakowatz}.

\begin{figure}[t]
\centering
\includegraphics[width=.7\columnwidth]{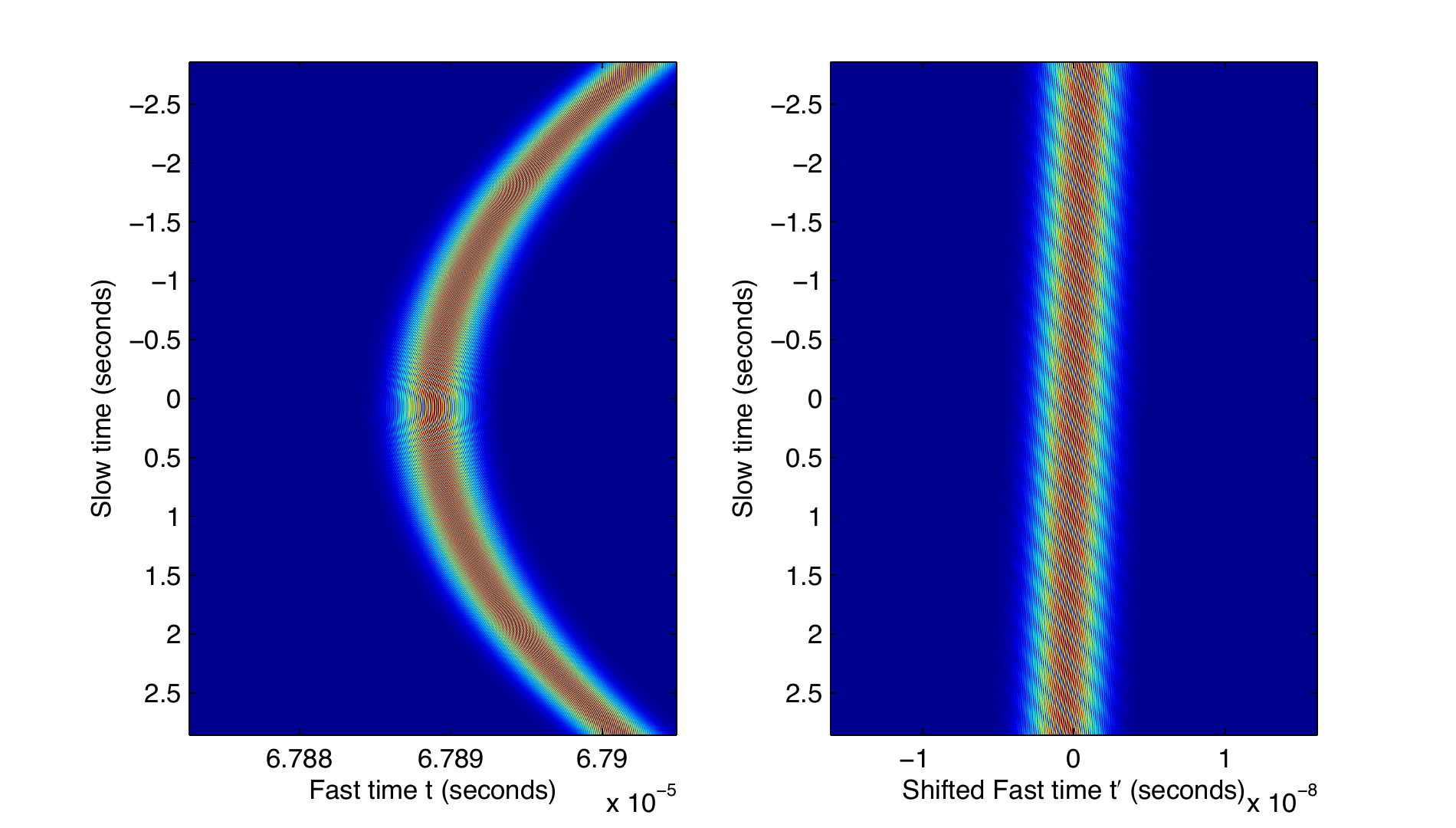}
\caption{Left: Pulse compressed data $D_p(s,t)=f(t-\tau(s,\vrho))$
  from a single scatterer at $\vrho=(0,5,0)$, recorded by a SAR
  antenna moving at speed $V=70$ m/s along a linear aperture of $400$
  m, at range $L = 10$ km from $\vrho$.  Note the hyperbolic curve
  defined by the peak.  Right: Pulse and range-compressed data $D_r(s,t')$ with
  respect to point $\vrho_o=(0,0,0)$. The abscissa is the shifted fast
  time $t'$. Note that most of the slow time dependence is
  removed in the range-compressed data.}
\label{fig:hyperbola}
\end{figure}
Another common data pre-processing step is \emph{range compression}.
It amounts to evaluating the pulse compressed data at times offset by
the round trip travel time to a reference point $\vrho_o \in
\cJ^{^\cI}$,
\begin{equation}
D_r(s,t') = D_p(s,t' +\tau(s,\vrho_o)).
\end{equation}
Here $t'$ is the fast time shifted by $\tau(s,\vrho_o)$, so that 
\[
t' + \tau(s,\vrho_o) = t \in [0,\Delta s].
\]
In the Fourier domain, the range compression amounts to removing the
large phase $\om \tau(s,\vrho_o)$ of $\hat D_p(s,\om)$.  This is
obviously advantageous from the computational point of view.  But
range compression plays a bigger role in our context. It is essential
for the robust PCA method to work.  To illustrate this, we show in
Figure \ref{fig:hyperbola} the pulse compressed echo from a single
scatterer, for a linear aperture. The scatterer is at $\vrho =
(0,5,0)$, with the first two coordinates defining the location in the
imaging plane and the third the elevation, which is always zero.  The
coordinates in the imaging plane are range and cross-range, with
origin at the reference point. The range is the coordinate of $\vrho$
along the direction pointing from the SAR platform (at the center of
the aperture) to $\vrho_o$. The cross-range is the coordinate of
$\vrho$ in the direction orthogonal to the range.

We plot in Figure \ref{fig:hyperbola} the amplitude of $D_p(s,t)$ as a
function of $s$ and $t$, and note that the location of the peak,
defined by equation
\[
\left(\frac{ct}{2}\right)^2 = |\vr(s)-\vrho|^2,
\]
is a hyperbola for the linear aperture. The peak lies on some other
curve in the $(s,t)$ plane for other apertures. The amplitude of the
range compressed data is shown in the right plot of Figure
\ref{fig:hyperbola}.  We note that the dependence on the slow time $s$
has been approximately removed by the range compression.  Explicitly,
it lies on the curve in the $(s,t')$ plane defined by equation
\[
\frac{ct'}{2} = |\vr(s)-\vrho| -|\vr(s)-\vrho_o|.
\]
This curve is close to the vertical axis $t'=0$ because
$\vrho$ and $\vrho_o$ are close to each other,
\[
|\vrho - \vrho_o| \ll |\vr(s)-\vrho|, \qquad \forall s.
\]
Consequently, the matrix with entries $D_r(s,t)$, sampled at discrete
$s$ and $t$, appears to be of low rank and it can be handled by the robust PCA
method.

We work with the pulse and range compressed data from now on, and to
simplify notation, we drop the prime from the shifted fast time $t'$.
We borrow terminology from the geophysics literature and call the
pulse and range compressed echoes \emph{data traces}.  The image is
formed by superposing the traces over the aperture, and
backpropagating them to the imaging points $\vrhoi \in \cJ^{^\cI}$
using travel times
\begin{eqnarray}
  \cI\left(\vrho^{\cI}\right)&=&\sum_{j = -n/2}^{n/2}  
  D_r(s_j,\tau(s_j,\vrhoi)-\tau(s_j,\vrho_o)) \nonumber \\
  &\approx& \frac{1}{\Delta s} \int_{-S(a)}^{S(a)} ds \,  
  D_r(s,\tau(s,\vrhoi)-\tau(s,\vrho_o)).
\label{eq:imagefxn}
\end{eqnarray}
Here $s_j$ are the discrete slow time samples in the interval
$[-S(a),S(a)]$ defining the aperture $a$ along the flight track.  The
sampling is uniform, at intervals $\Delta s$, and
\[
2 S(a) = n \Delta s, \quad \mbox{with} ~ n ~ \mbox{even}.
\]
Assuming a large $n$, that is a small $\Delta s$, we approximate the
sum in (\ref{eq:imagefxn}) by an integral over the aperture.

When the aperture is very large, the data in (\ref{eq:imagefxn}) is
weighted by a factor that compensates for geometrical spreading
effects over the long flight track, and thus improves the focus of the
image.  Here we work with small apertures where geometrical spreading
plays no role, which is why there are no weights in the imaging
function (\ref{eq:imagefxn}).

\section{Robust PCA for motion detection and SAR data separation}
\label{sect:decomp}
\setcounter{equation}{0}

We begin in section \ref{sect:RPCA} with a brief discussion of the
robust PCA method. Then, we give in section \ref{sect:explain} a
heuristic explanation of why it makes sense to use it for motion
detection and data separation. We also illustrate in section
\ref{sect:rpca} the difficulties arising in the separation, and the
improvements achieved by proper data windowing.

\subsection{Robust PCA}
\label{sect:RPCA}
The robust PCA method, introduced and analyzed in \cite{candesRPCA},
applies to matrices $M \in \mathbb{R}^{n_1 \times n_2}$ that are sums of a
low rank matrix $\cL_o$ and a sparse matrix $\cS_o$. It solves a convex
optimization problem called principle component pursuit:
\begin{align}
  & \min_{\cL,\cS\in\mathbb R^{n_1\times n_2}}
  \quad ||\cL||_*+ \eta ||\cS||_1 \\
  & \text{subject to} \quad \cL+\cS=M, 
\end{align}
with 
\begin{equation}
\eta = \frac{1}{\sqrt{\max
      \{n_1, n_2\}}}.
\end{equation}
Here $||\cL||_*$ is the nuclear norm, i.e. the sum of the
singular values of $\cL$, and $||\cS||_1$ is the matrix 1-norm
of $\cS$.  The optimization can be done for any matrix, but the point is
that if $M = \cL_o + \cS_o$, with $\cL_o$ low rank and
$S_o$ sparse and high rank, then the principle component pursuit
recovers exactly $\cL_o$ and $\cS_o$.
The analysis in \cite{candesRPCA} gives sufficient conditions under
which the decomposition is exact. These conditions are bounds on the
rank of $\cL_o$ and the number of non-zero entries in the high rank
matrix $\cS_o$. They are not necessary conditions, meaning that the
decomposition can be achieved for a much larger class of matrices than
those fulfilling the assumptions of the theorems in
\cite{candesRPCA}. This is already pointed out in \cite{candesRPCA}.

\subsection{The structure of the matrix of range compressed SAR data}
\label{sect:explain}
We illustrate in this section the pulse and range compressed echoes
(the traces) from a complex scene.  The point of the illustration is
to show that typically, the sampled traces from the stationary scene
form a low rank matrix, whereas those from moving targets give a high
rank but sparse matrix.  We refer to section \ref{sect:numerics} for
the description of the setup of the numerical simulations used to
produce the results presented here and in the next section.

\begin{figure}[t]
\hspace{-.7in}
\includegraphics[width=1.2\columnwidth]{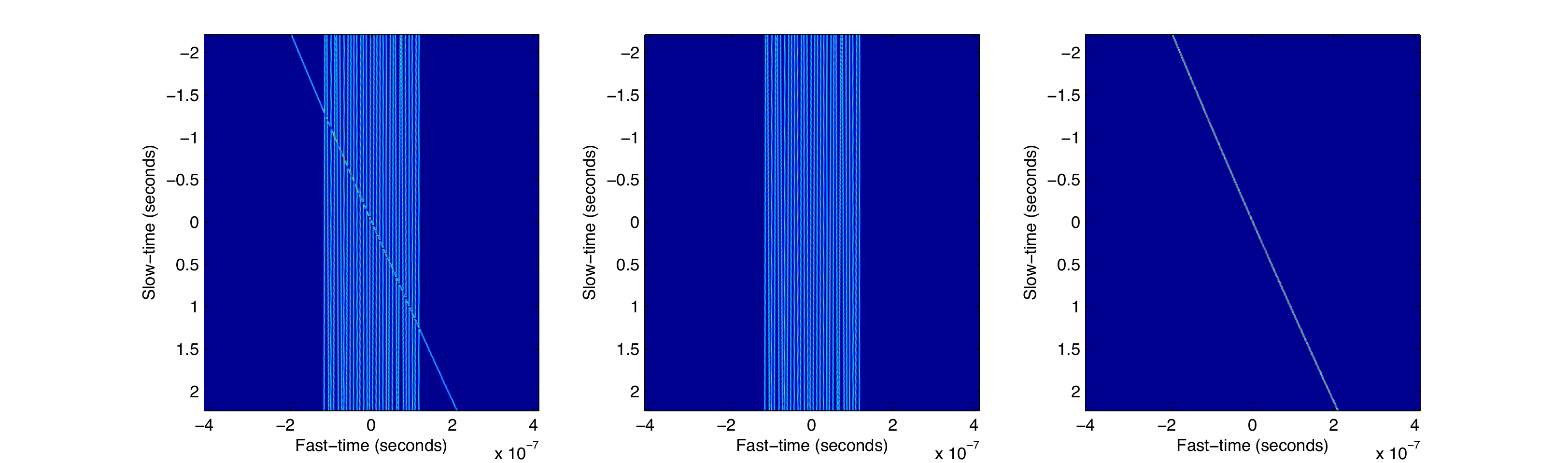}
\caption{Desired separation of the range and pulse compressed SAR
  data.  Left: $|D_r(s,t)|$ for a sample scene with thirty stationary
  targets.  The vertical traces correspond to the stationary targets,
  while the sloped trace corresponds to the moving target.  Middle:
  The stationary component of $|D_r(s,t)|$.  Right: The moving
  component of $|D_r(s,t)|$.  We plot the absolute value of these matrices 
  only to increase contrast for the figures.}
\label{fig:data}
\end{figure}

Let $M$ be the matrix with entries given by the data traces sampled at
the discrete slow and fast times
\begin{equation}
M_{jl}=D_r(s_j,t_l), 
\label{eq:defM1}
\end{equation}
where 
\begin{equation}
s_j = j \Delta s, \quad j = -n/2, \ldots, n/2,
\label{eq:defM2}
\end{equation}
and 
\begin{equation}
t_l = l \Delta t, \quad l = -m/2,\ldots, m/2.
\label{eq:defM3}
\end{equation}
The sampling is at uniform intervals $\Delta s$ and $\Delta t$, and $n
$ and $m $ are positive integers satisfying
\begin{equation}
2 S = n \Delta s, \quad \mbox{and} \quad 
\Delta s = m \Delta t.
\label{eq:defM4}
\end{equation}
We assume throughout that $n$ and $m$ are even and large.

We show in the left plot of Figure \ref{fig:data} the matrix $M$ for a
scene with a single moving target and thirty stationary
scatterers. The ideal separation of this matrix would be
\[
M = \cL_o + \cS_o
\]
where $\cL_o$ is given by the echoes from the stationary targets alone
and $\cS_o$ is given by the echo from the moving target. We plot
$\cL_o$ and $\cS_o$ in the middle and right plot of Figure
\ref{fig:data}.  It appears to the eye that $\cL_o$ is a matrix with
almost parallel columns, so we expect it to be low rank. The matrix
$\cS_o$ is given by the sloped curve in the $(s,t)$ plane. The range
compression does not remove the $s$ dependence of the echo from the
moving target, as is the case for the stationary ones, and this is why
we see this sloped curve. Consequently, $\cS_o$ has higher rank than
$\cL_o$. But it is sparse, because we have only one moving target,
i.e., one sloped curve. Obviously, the matrix $\cS_o$ will remain
sparse for a small number of moving targets, as well.

Thus, the matrix $M$ appears to have the appropriate structure for the
robust PCA method to be useful. But we show next that robust PCA alone
will not do the job adequately. It must be complemented with proper
data windowing.

\subsection{Data windowing for separation with robust PCA}
\label{sect:rpca}
\setcounter{equation}{0} We illustrate with two numerical simulations
that robust PCA cannot be applied as a black box to the SAR data
traces and produce the desired separation of the stationary and
moving target parts.  However, if it is applied to properly calibrated
subsets (or {\em windows}) of the data, robust PCA can separate the
traces for a large class of complex scenes.

\begin{figure}[t]
 \subfigure{\includegraphics[width=.36\columnwidth]{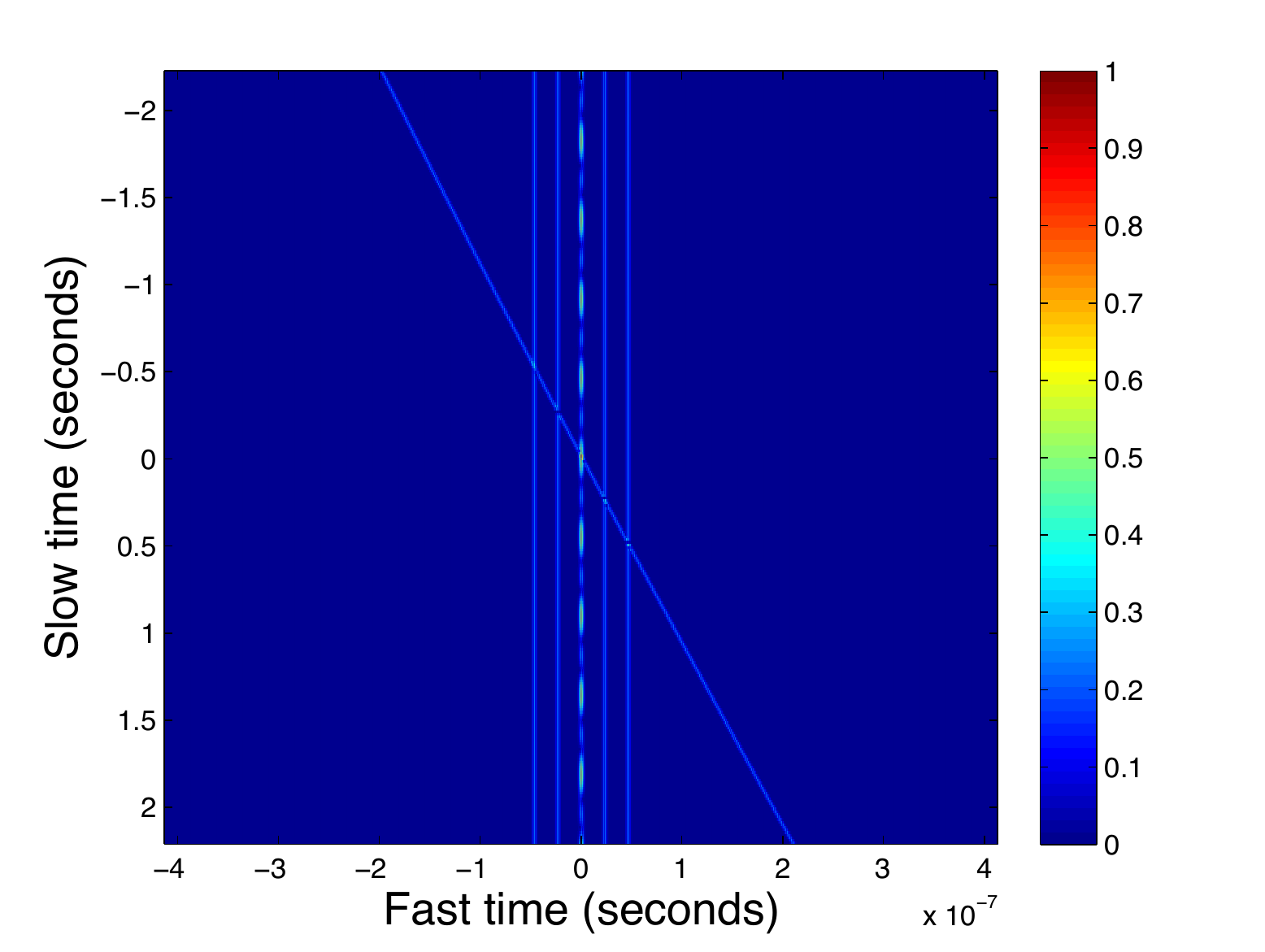}}\hspace{-0.3in}
 \subfigure{\includegraphics[width=.36\columnwidth]{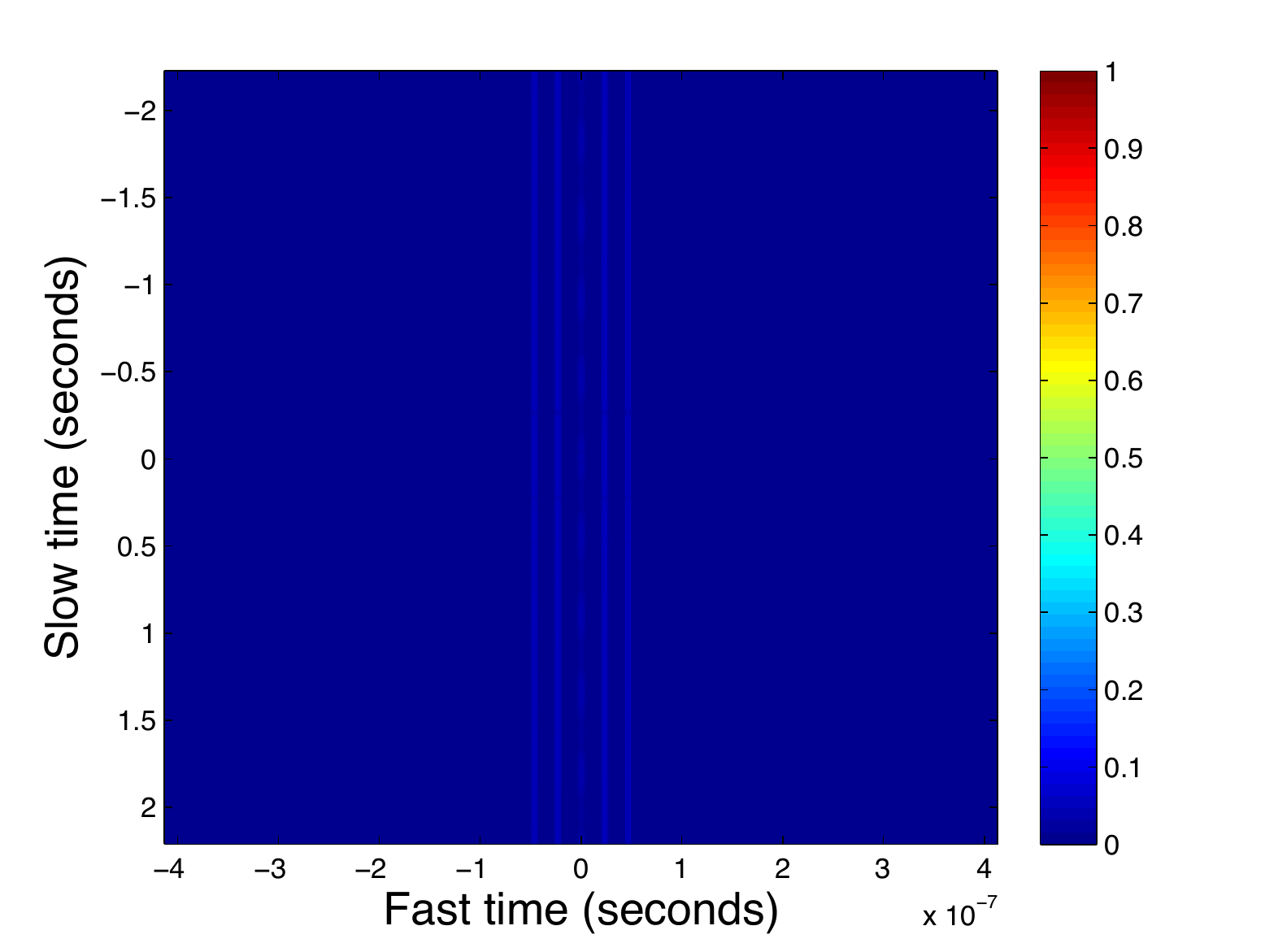}}\hspace{-0.3in}
 \subfigure{\includegraphics[width=.36\columnwidth]{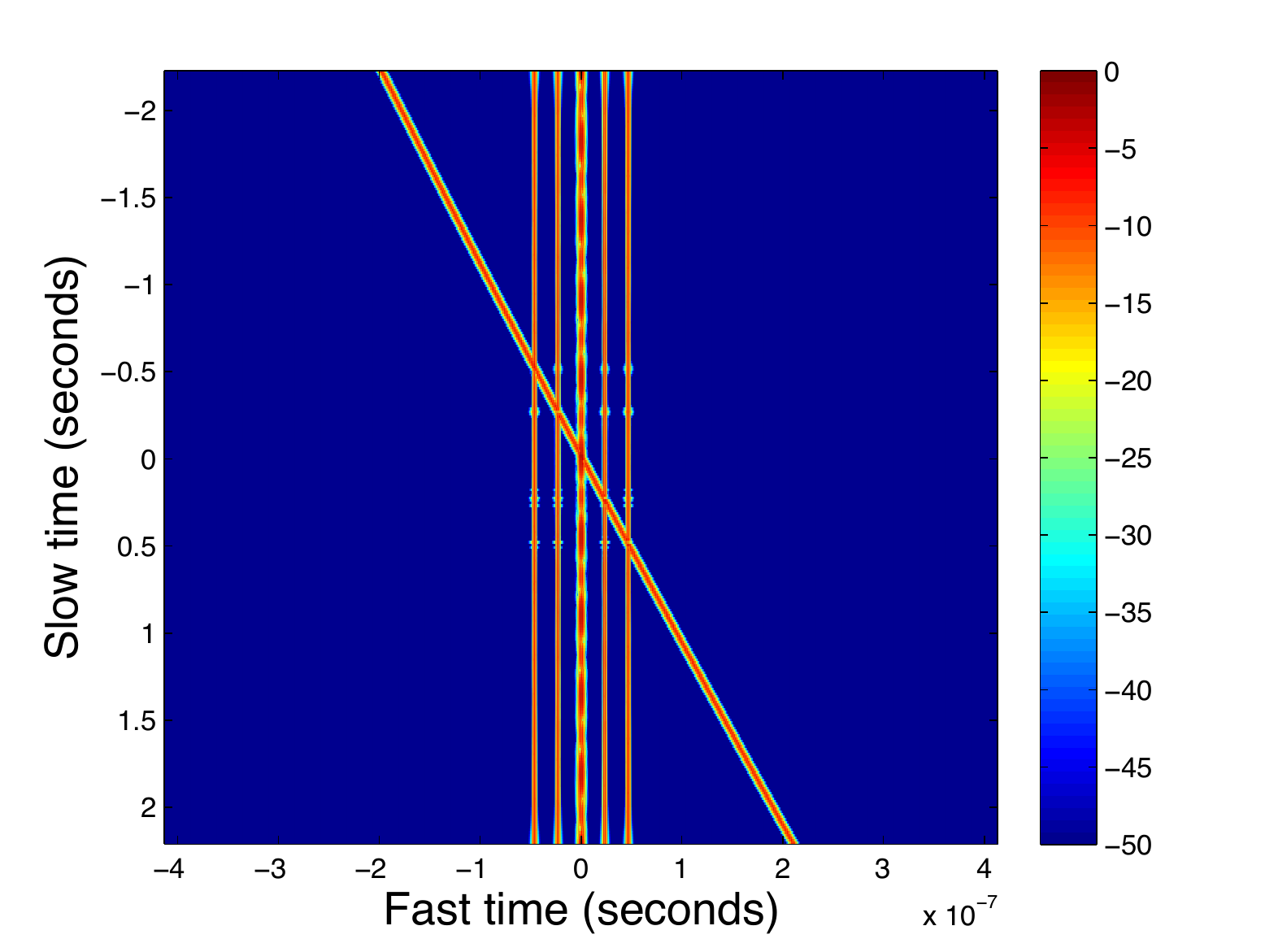}}\\
 \subfigure{\includegraphics[width=.36\columnwidth]{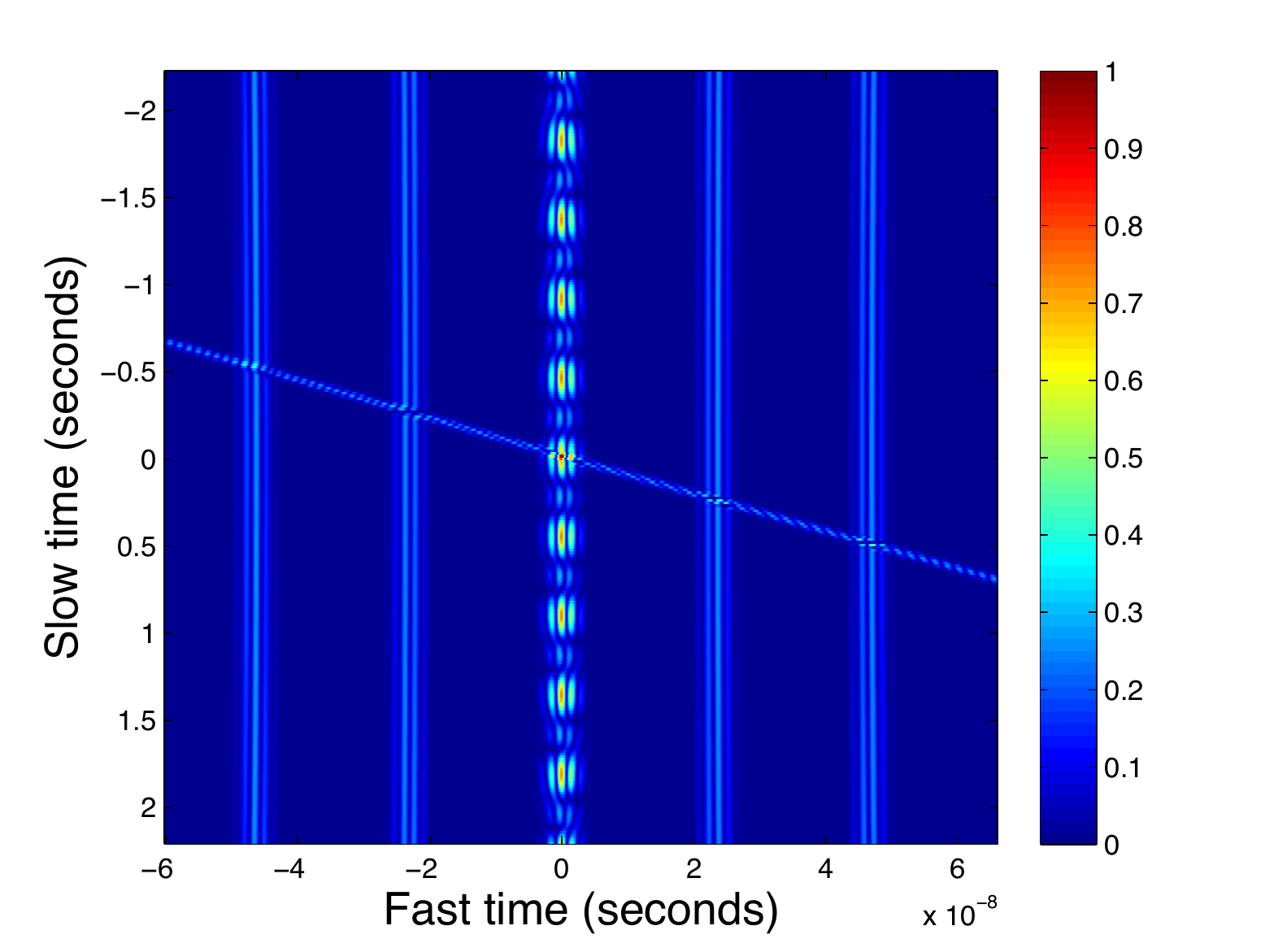}}\hspace{-0.3in}
 \subfigure{\includegraphics[width=.36\columnwidth]{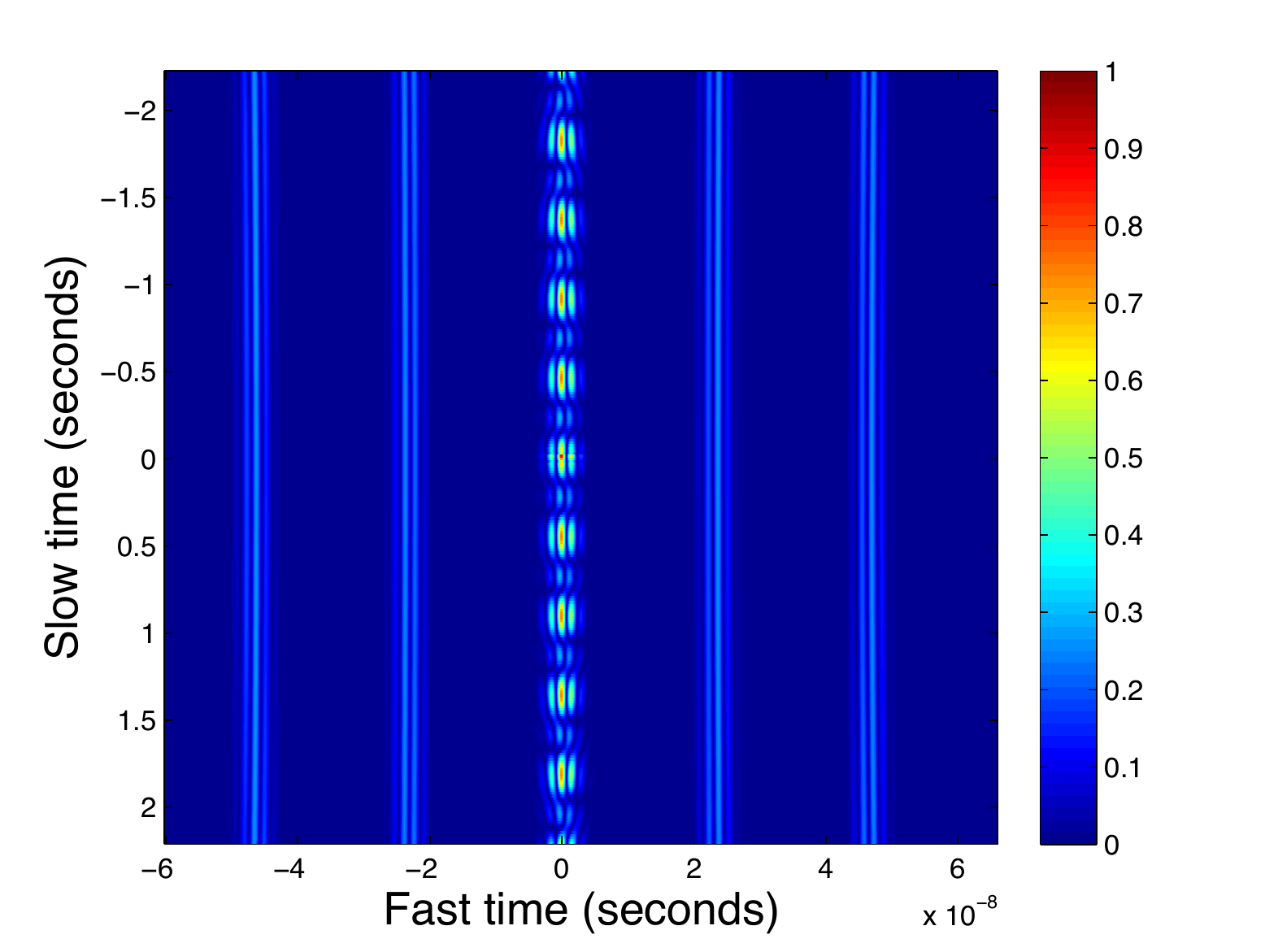}}\hspace{-0.3in}
 \subfigure{\includegraphics[width=.36\columnwidth]{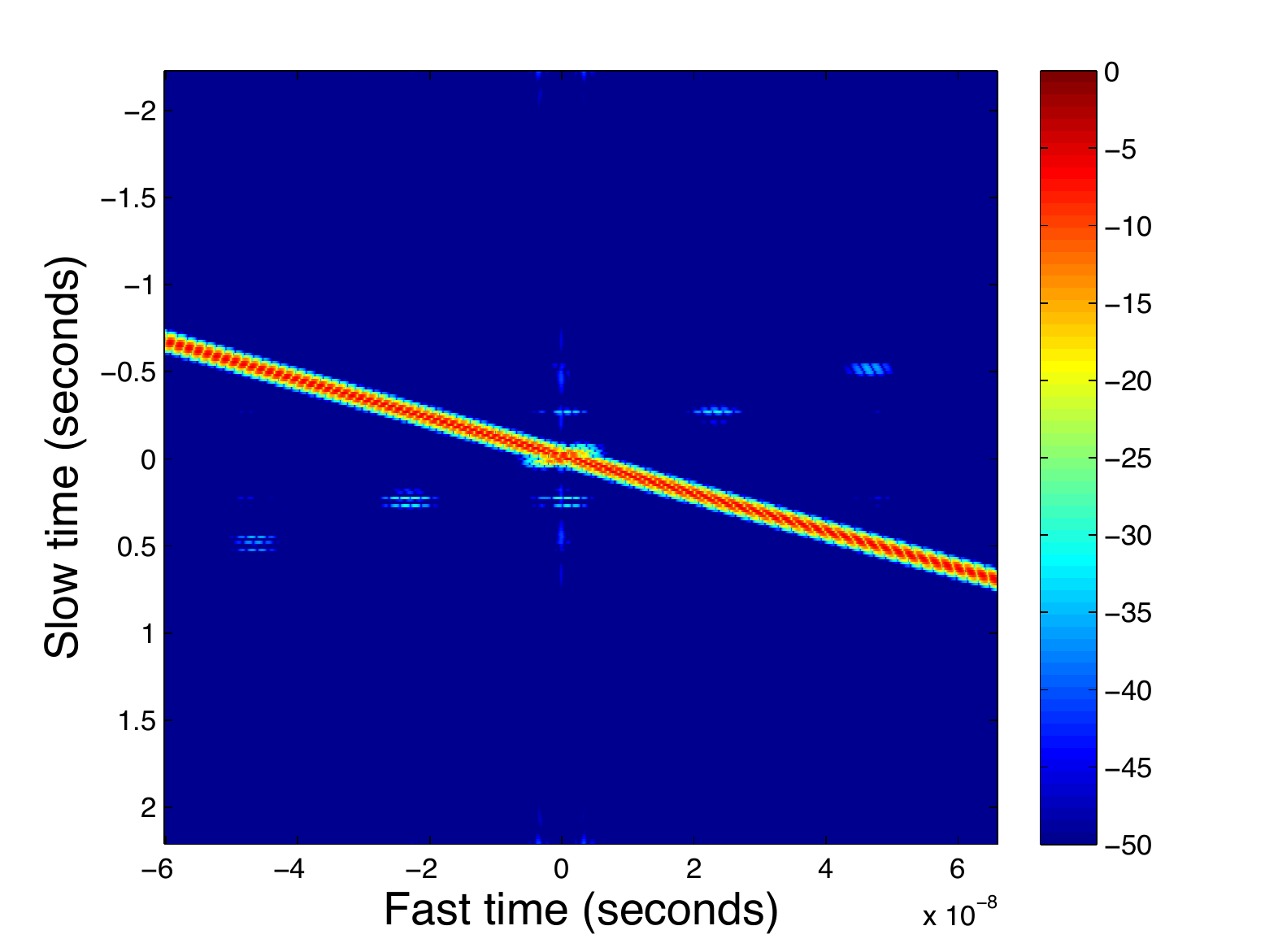}}
\caption{Robust PCA separation of traces from a scene with seven
  stationary targets and a moving target.  From left to right we show
  the matrix $M$ of traces, the low rank $\cL$, and sparse $\cS$
  components from robust PCA.  Top row is robust PCA applied to the
  entire data, bottom row is robust PCA applied to a windowed part of
  the data.  In each plot the matrices are normalized by the largest
  value of $|D_r(s,t)|$.  The sparse components shown in the right
  column are plotted in (decibel) dB scale to make the contrast more
  visible.  }
\label{fig:rpcaEx1}
\end{figure}

The results of the first simulation are in Figure \ref{fig:rpcaEx1}.
We have a scene with seven stationary targets at
  coordinates $(0,0,0)$m, $(\pm 5,0,0)$m, $(0,\pm 5,0)$m and $(\pm 10,0,0)$m in the
  imaging plane, which is at elevation zero. There is also a moving
target that is at location $(0,0,0)$, at slow time
$s=0$ corresponding to the center of the aperture. The target moves
with velocity ${\bf u} = \frac{28}{\sqrt{2}} (1,1)$m/s in the plane.  We
refer to section \ref{sect:numerics} for a detailed description of the
setup for the simulations. In the top row of plots in Figure
\ref{fig:rpcaEx1} we show the results with robust PCA applied to the
matrix $M$ of data traces shown on the left.  The middle and right
plots are the resulting low rank and sparse parts given by the robust
PCA algorithm.  Due to the fact that the entire matrix $M$ is sparse
to begin with, much of the stationary target data is captured by the
sparse component. Thus, we do not have a good separation.  However,
the result is very good when we apply the robust PCA algorithm on a
smaller fast-time window of the data, as shown in the second row of
plots.  The number of nonzero entries in the windowed $M$ is no longer
small relative to its size, and robust PCA separates the moving target
trace from the rest.

\begin{figure}[t]
\centering
\includegraphics[width=.5\columnwidth]{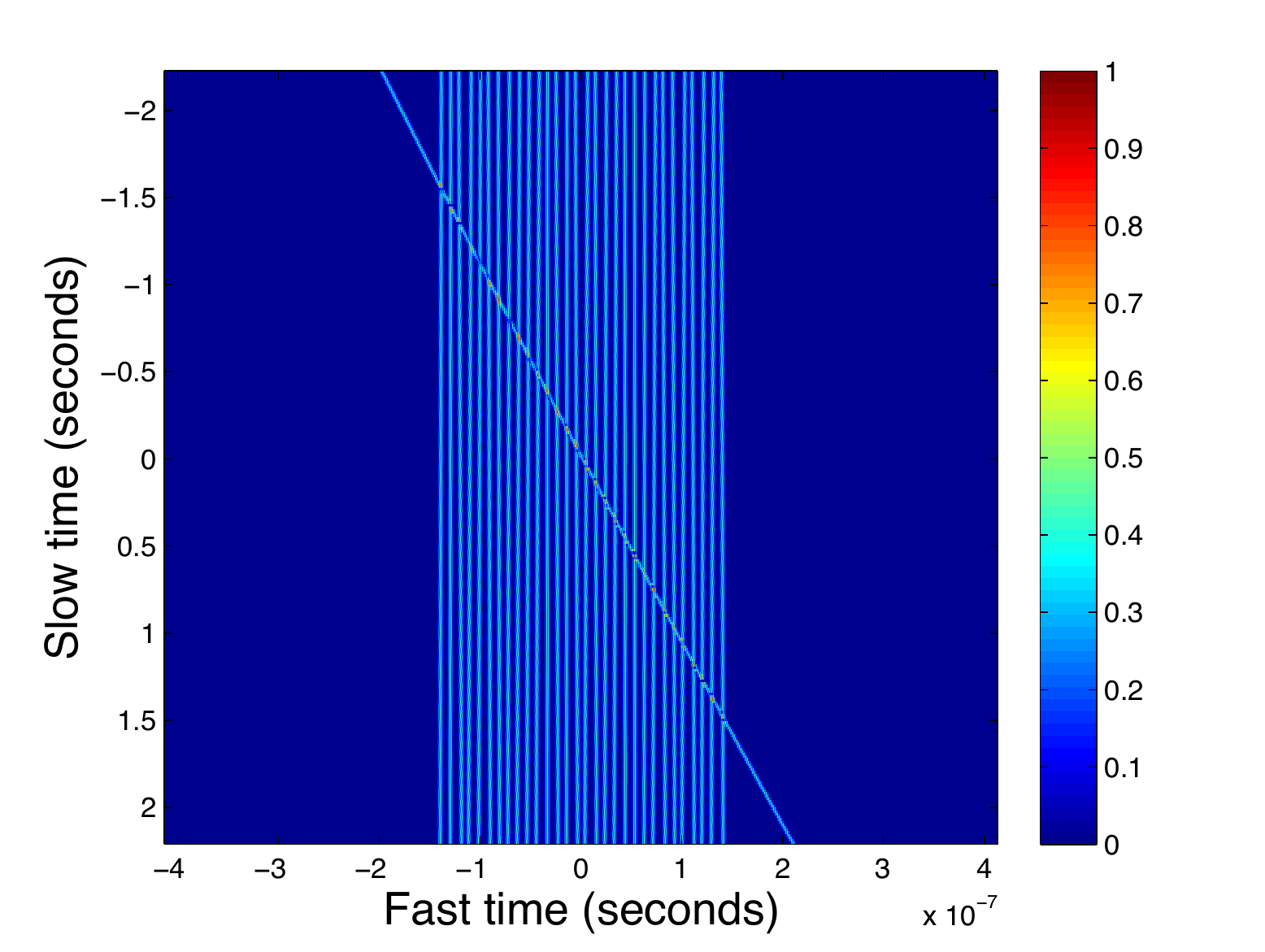}
\caption{SAR scene with thirty stationary targets and a moving one.}
\label{fig:rpcaEx2a}
\end{figure}

\begin{figure}[t]
\centering
\includegraphics[width=.99\columnwidth]{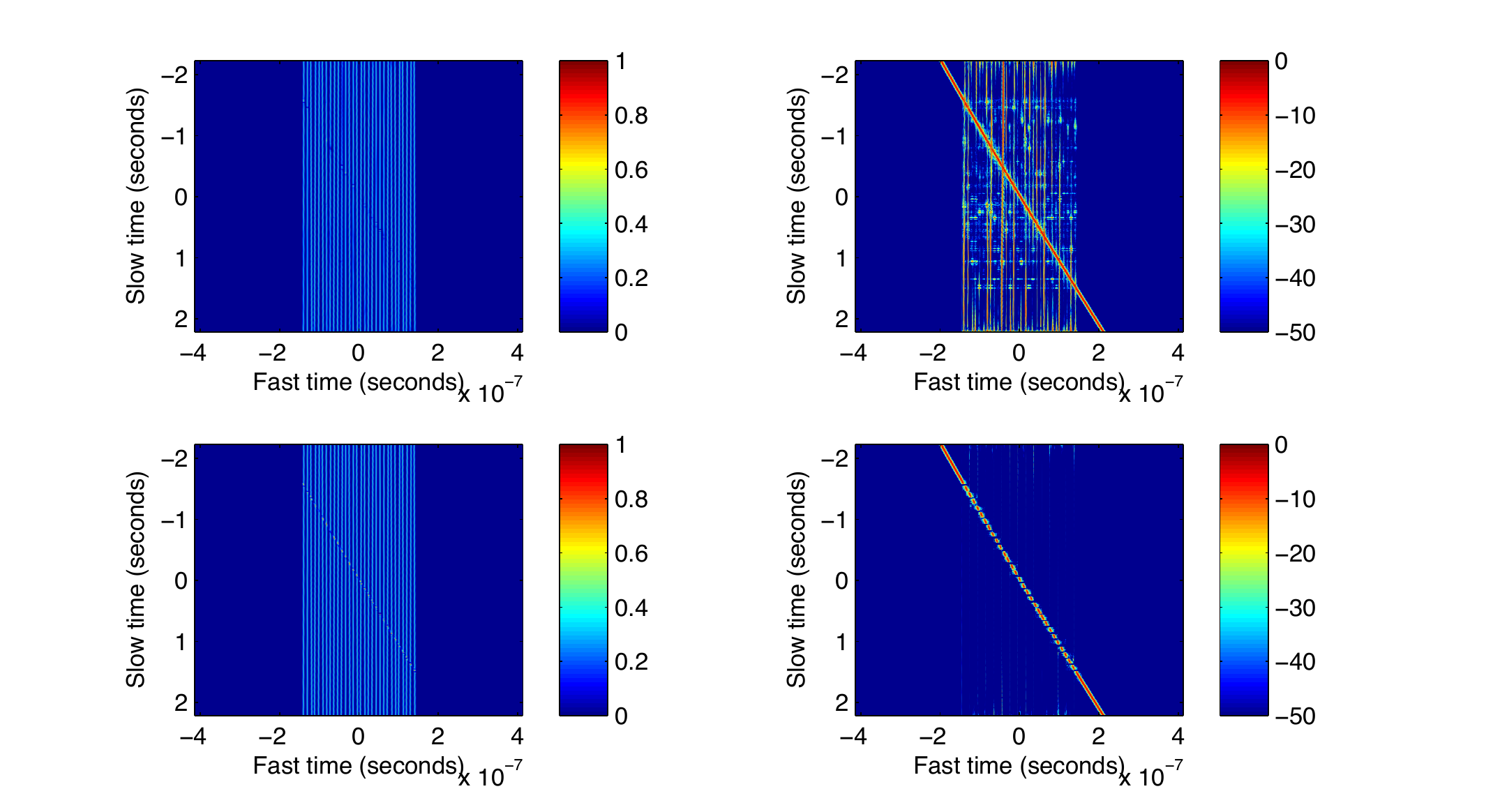}
\caption{Robust PCA separation of traces from a scene with thirty
  stationary targets and a moving one.  The top row of plots shows the
  results with robust PCA applied to the entire data matrix
  in Figure \ref{fig:rpcaEx2a}. The low rank and sparse components are shown in the
  middle and on the right.  The bottom row of plots shows the
  results with robust PCA applied to successive small time
  windows of the data.  The sparse components shown in the right
  column are plotted in dB scale to make the contrast more visible.}
\label{fig:rpcaEx2b}
\end{figure}

In Figure \ref{fig:rpcaEx2b} we show the results for a scene with
thirty stationary targets and a moving one. The data traces from this
scene are displayed in Figure \ref{fig:rpcaEx2a}. The moving target is
as in the previous example. The point of the simulation is to show
that when we work in a large fast time window the traces from all the
stationary targets may no longer form a low rank matrix. The top row
of plots in Figure \ref{fig:rpcaEx2b} shows that the sparse component
of the matrix $M$, as returned by the robust PCA algorithm, has a
large residual part from the stationary targets.  The much improved
results in the bottom are obtained by applying robust PCA on
successive small fast time windows of the data, and then reassembling
the separated traces. To give an idea of the size of the windows, the
matrix $M$ has dimensions $n = 296$ by $m = 16,384$. The matrices are windowed
in fast-time and are of size $296 \times 450$.

The two examples given above show that data windowing plays an
important role in achieving a successful data separation with robust
PCA. The last simulation also illustrates the role of robust PCA in
the detection of the moving target. While the trace from this target
is faint and difficult to distinguish from the others over most of the
fast time interval in Figure \ref{fig:rpcaEx2a}, it is clearly visible
in Figure \ref{fig:rpcaEx2b} after the processing of the traces with
robust PCA.

\section{Analysis of rank of the matrix of SAR data traces}
\label{sect:rank}
\setcounter{equation}{0} We present here an analysis of the rank of
the matrix of SAR data traces for simple scenes with one or two
targets. The goal is to understand how the rank depends on the
position of the targets and their velocity. We limit the analysis to
at most two targets to get a simple structure of the matrix
$M M^T$, for which we can calculate the rank almost explicitly. The
numerical results presented above and in section \ref{sect:numerics}
show that the data separation works for complex scenes, with many
stationary targets. 

We begin in section \ref{sect:scaling} with the scaling regime used in
the analysis. We illustrate it with the GOTCHA Volumetric SAR data set
\cite{Gotcha} for X-band surveillance SAR. The model of the matrix $M$
is given in section \ref{sect:model}.  We analyze its rank in section
\ref{sect:single} for a single target, and in section
\ref{sect:double} for two targets. We end with a brief discussion in 
section \ref{sect:rankDisc}.

\subsection{Scaling regime and illustration with GOTCHA volumetric
  SAR}
\label{sect:scaling}
The important scales in the problem are: The central frequency
$\nu_o$, the bandwidth $B$, the typical distance (range) $L$ from the
SAR platform to the targets, the aperture $a$, the magnitude $|\bu|$
of the speed of the targets, the speed $V$ of the SAR platform and
$R^{\cI}$, the diameter of the imaging set $\cJ^{^\cI}$. We assume
that they are ordered as follows
\begin{equation}
B \ll \nu_o,
\label{eq:SC1}
\end{equation}
and 
\begin{equation}
a \ll L, \quad R^{^\cI} \ll L,
\label{eq:SC2}
\end{equation}
and we let 
\begin{equation}
L = |\vr(0) - \vrho_o|.
\end{equation}
We also suppose that the target speed is smaller than that of the SAR
platform
\begin{equation}
|\bu| < V.
\label{eq:SC3}
\end{equation}

As an illustration, consider the setup in GOTCHA Volumetric SAR.  The
central frequency of the probing signal is $\nu_0=9.6$GHz and
the bandwidth is $B=622$MHz, so assumption (\ref{eq:SC1}) holds.  The
SAR platform trajectory is circular, at height $H=7.3$km, with radius
$R = 7.1$km and speed $V=250$km/h or $70$m/s. One circular degree of
trajectory is $124$m.  The pulse repetition rate is $117$ per degree,
which means that a pulse is sent every $1.05$m, and $\Delta s =
0.015$s.  A typical distance to a target is $L=10$km and we consider
imaging domains of radius $R^{^\cI}$ of at most $50$m, so assumptions
(\ref{eq:SC2}) hold.  The target speed is $|\bu| \sim 100$km/h or
$28$m/s, so it satisfies (\ref{eq:SC3}).  

For a stationary target we obtain from basic resolution theory 
(for single small aperture imaging) that
the range can be estimated with precision $c/B=48$cm, and the cross
range resolution is $\lambda_0 L/a = 2.5$m, with one degree aperture
$a$ and central wavelength $\la_0 = 3$cm. The image of a moving target 
is out of focus unless we estimate its velocity and compensate 
for the motion in the imaging function.

\subsection{Data model}
\label{sect:model}
Our model of the data assumes that the scatterers lying on
the imaging surface behave like point targets. We neglect any
interaction between the scatterers, meaning that we make the single
scattering, Born approximation. The pulse and range compressed data 
is approximated by 
\begin{equation}
  D_r(s,t)\approx \sum_{q = 1}^{N}\frac{\sigma_q(\om_o)}{
    (4\pi|\vr(s)-\vrho_q(s)|)^2}f_p(t-(\tau(s,\vrho_q(s))-
  \tau(s,\vrho_o))),
\label{eq:MOD1}
\end{equation}
in the case of $N$ targets at locations 
\[
\vrho_q(s) = (\brho_q(s),0), \quad q = 1, \ldots, N,
\]
with reflectivity $\sigma_q(\om_o)$.  Here we used the so-called
stop-start approximation which neglects the displacement of the
targets during the round trip travel time. This is justified in radar
because the waves travel at the speed of light that is many orders of
magnitude larger than the speed of the targets. We refer to \cite{sar}
for a derivation of the model (\ref{eq:MOD1}) in the scaling regime
described above.

Because of our scaling assumptions (\ref{eq:SC2}), we can approximate
the amplitude factors as
\[
\frac{1}{4 \pi |\vr(s)-\vrho_q(s)|} \approx \frac{1}{4 \pi L},
\]
and obtain the simpler model 
\begin{equation}
\label{eq:MOD2}
D_r(s,t)=\left(\frac{1}{4\pi L}\right)^2 \sum_{q=1}^N
\sigma_q f_p(t-\Delta\tau(s,\vrho_q(s))),
\end{equation}
where we let 
\begin{equation}
\label{eq:MOD3}
\Delta\tau(s,\vrho(s))=\tau(s,\vrho(s))-\tau(s,\vrho_0).
\end{equation}  

We assume henceforth, for simplicity, that the targets are identical
\[
\sigma_q = \sigma, \quad q = 1, \ldots, N,
\]
and write 
\begin{equation}
D_r(s,t) \approx \frac{\sigma(\om_o)}{(4 \pi L)^2} {\mathcal M}(s,t),
\label{eq:MOD4}
\end{equation}
with 
\begin{equation}
{\mathcal M}(s,t) = \sum_{q=1}^N
f_p(t-\Delta\tau(s,\vrho_q(s))).
\label{eq:MOD5}
\end{equation}
The matrix of traces analyzed below is given by discrete 
samples of (\ref{eq:MOD5}),
\begin{equation}
  M_{jl} = {\mathcal M}\left(s_{j-\frac{n}{2}-1},
    t_{l-1}\right), \quad j = 1, \ldots, n+1, ~ ~ l = 1, 
  \ldots, m + 1,
\label{eq:MOD6}
\end{equation}
with slow times $s_j$ and fast times $t_l$ defined in (\ref{eq:defM2})
and (\ref{eq:defM3}). We also take for convenience a compressed pulse
given by a Gaussian modulated by a cosine at the central frequency,
\begin{equation}
f_p(t) = \cos(\om_o t) e^{-B^2 t^2/2}.
\label{eq:MOD7}
\end{equation}

\subsection{Analysis of rank of the data traces for one target}
\label{sect:single}
In the case of one target at location $\vrho(s) = (\brho(s),0)$, the
entries of matrix (\ref{eq:MOD6}) are given by
\begin{equation}
  M_{jl} = \cos\left[\om_o(t - \Delta\tau(s,\vrho(s)))\right]
  \exp \left[ -\frac{B^2}{2} (t - \Delta\tau(s,\vrho(s)))^2 \right],
\label{eq:MOD8}
\end{equation}
with $\Delta \tau$ defined by (\ref{eq:MOD3}). 
The target is moving at speed $\vec{\bf u} = (\bu,0)$, so we have that 
\begin{equation}
\vrho(s) = \vrho + s \, \vec{\bu}, \quad |s| \le S(a),
\label{eq:MOD9}
\end{equation}
where we let
\[
\vrho := \vrho(0)
\]
be the location of the target at the time $s = 0$, corresponding to
the center of the aperture.

We study the rank of $M$, which is equivalent to studying the rank of
the symmetric, square matrix $C \in \mathbb{R}^{(n+1)\times (n+1)}$,
with entries given by
\begin{equation}
  C_{j,l} = \cC\left(s_{j-\frac{n}{2}-1},
    s_{l-\frac{n}{2}-1}\right), \qquad 
  j,l = 1, \ldots, n+1,
\label{eq:covar1}
\end{equation}
in terms of the function 
\begin{equation}
  \cC(s,s') =  \sum_{q=-m/2}^{m/2} D_r(s,t_q) D_r(s',t_q)
\approx \frac{1}{\Delta t}\int_{-\infty}^\infty dt\ D_r(s,t)D_r(s,t).
\label{eq:covar1p}
\end{equation}
Here we used the assumption that $\Delta t$ is small enough to
approximate the Riemann sum over $q$ by the integral over $t$. Because
the traces $D_r(s,t)$ vanish for $|t| > \Delta s/2$, we extended the
integral to the whole real line.

We obtain after a calculation given in appendix \ref{sect:appSingle}
that if the aperture $a$ is small enough, the matrix $C$ has a
Toeplitz structure.
\begin{proposition}
\label{prop:Toep}
  Assuming that the Fresnel number $\frac{a^2}{\la_o L}$ is bounded by
\begin{equation}
  \frac{a^2}{\la_o L} \ll \min \left\{ 
    \frac{L}{R^{^\cI}}, \frac{V}{|\vu|} \right\},
\label{eq:Fresnel}
\end{equation}
the matrix $C$ is approximately Toeplitz, with entries given by 
\begin{equation}
  \cC(s,s') \approx \frac{\sqrt{\pi}}{2 B \Delta t} 
    \cos \left[ \om_o \alpha (s-s')\right] \exp \left[-\frac{(B \alpha)^2(s-s')^2}{4} \right].
\label{eq:Toep1}
\end{equation} 
The dimensionless parameter $\alpha$ depends linearly on the velocity
in the range direction and the cross-range offset of the target with
respect to the reference point $\vrho_o$. It is given by
\begin{equation}
  \alpha =\frac{2 \vu \cdot \vm_o}{c}-\frac{2V \vt \cdot \Pp_o(
\vrho-\vrho_o)}{
    cL}+\frac{2 \vu \cdot \Pp_o(\vrho-\vrho_o)}{cL},
\label{eq:Toep2}
\end{equation}
with $\vm_o$ the unit vector pointing in the range direction from the
center $\vr(0)$ of the aperture
\begin{equation}
\vm_o = \frac{\vr(0)-\vrho_o}{|\vr(0)-\vrho_o|},
\end{equation}
and $\Pp_o$ the orthogonal projection 
\begin{equation}
\Pp_o = I - \vm_o \vm_o^T.
\end{equation}
The unit vector $\vt$ is defined by  
\begin{equation}
\frac{d \vr(0)}{ds}  = V \vt.
\end{equation} 
It is tangential to the flight track, at the center
of the aperture.
\end{proposition}

\begin{figure}[!h]
\centering
\includegraphics[width=.4\columnwidth]{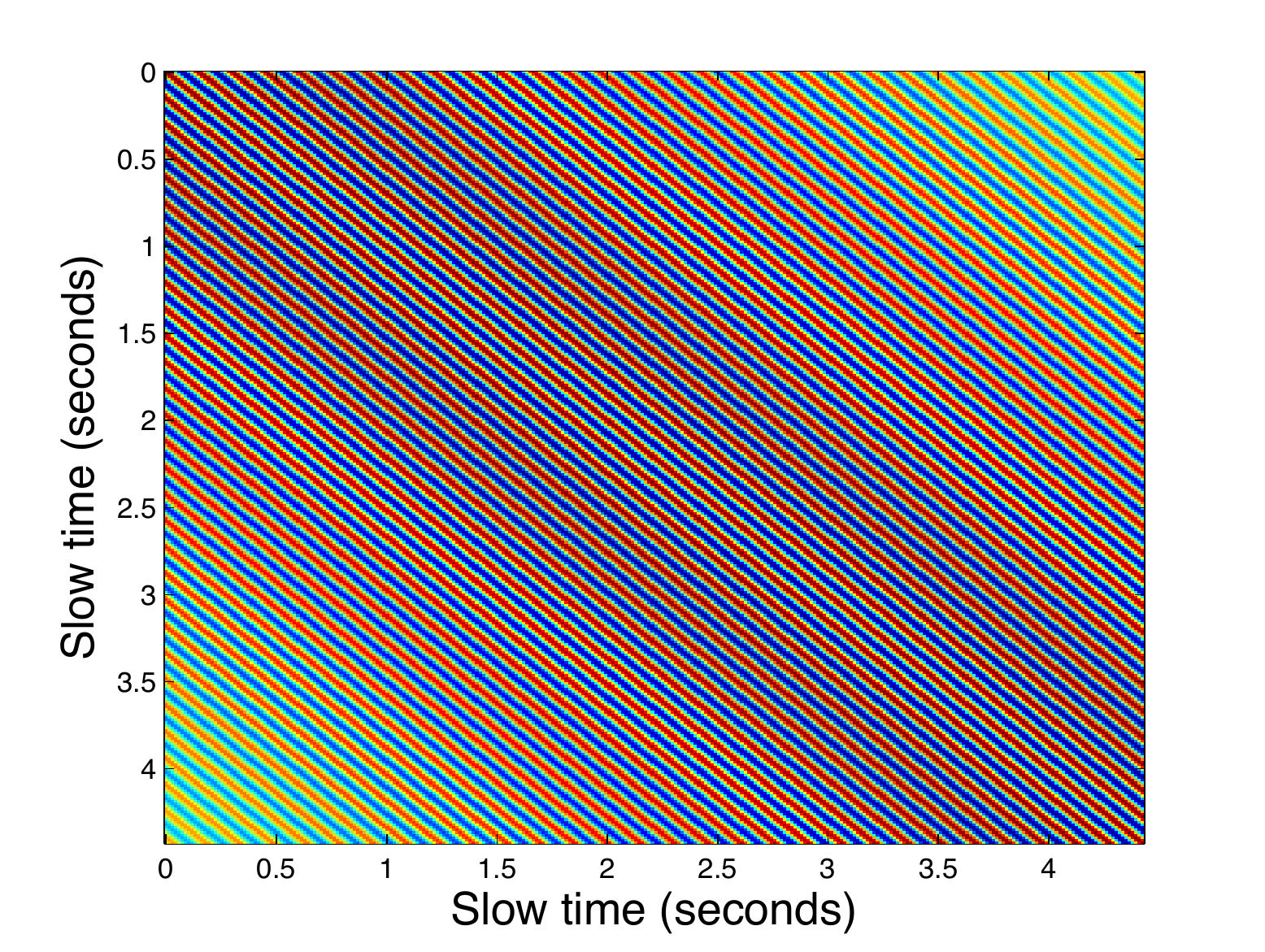}
\caption{The matrix $C$ for one stationary target at $\vrho=(0,15,0)$m.
The reference point is $\vrho_o= \vec{\bf 0}$. The aperture is $310$m, 
the range $L = 10$km and the central wavelength is $\la_o = 3$cm. The plot 
shows that the matrix is essentially constant along the diagonals, it is 
approximately Toeplitz.}
\label{fig:toeplitz1}
\end{figure} 
As an illustration, we plot in Figure \ref{fig:toeplitz1} the matrix
$C$ for an aperture $a = 310$m, in the GOTCHA regime with central
wavelength $\la_o = 3$cm and range $L = 10$km. We have a stationary
target at $\vrho = (0,15,0)$m, and $\vrho_o$ is at the origin. The
Fresnel number
\begin{equation}
\frac{a^2}{\la_o L} = 320.3 
\end{equation}
is only half the ratio $L/R^{^\cI}$, with $R^{^I} = 15$m, and yet the
matrix $C$ is essentially Toeplitz.

\begin{figure}[t]
\centering\includegraphics[width=.6\columnwidth]{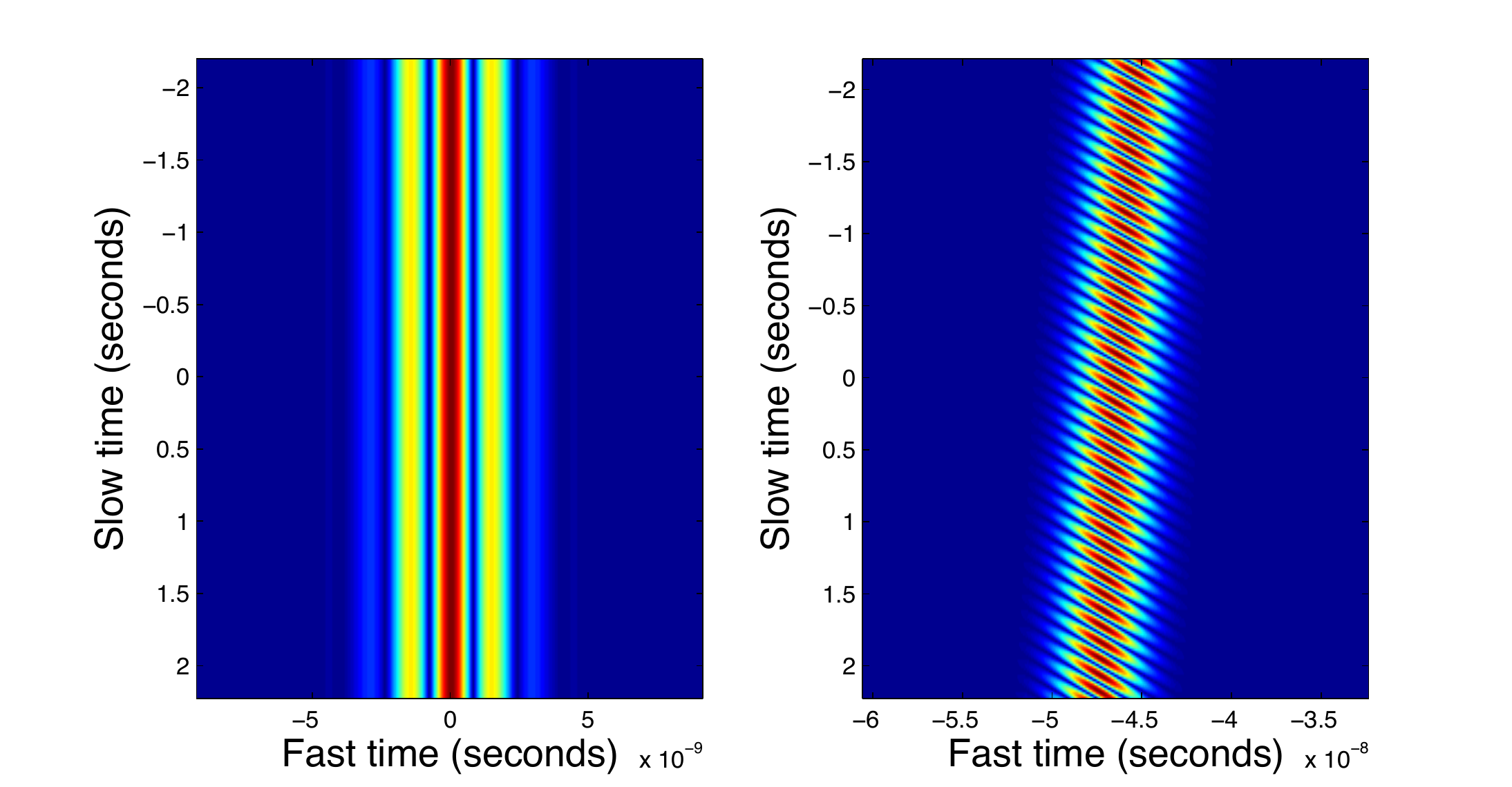}
\caption{Matrix $M$ of data traces for a stationary target located at
  $\vrho = \vrho_o=(0,0,0)$m (left) and at $\vrho = (-10,-10,0)$m
  (right).}
\label{fig:traces}
\end{figure}

\vspace{0.1in}
\noindent \textbf{Remark:} Since \[
\|D_r(s,\cdot)\|^2_2=\frac{\sqrt{\pi}}{2B\Delta t}, \quad \forall s,
\] 
the angle between two rows of the matrix $M$ of data traces, indexed
by $s$ and $s'$, is given by
\begin{equation}
\label{eq:ANGLE}
\cos \angle(D_r(s\cdot),D_r(s',\cdot))\approx\cos[\omega_o\alpha
(s-s')]\exp\left[-\frac{(B\alpha)^2 (s-s')^2}{4}\right].
\end{equation}
Nearby rows with indices satisfying 
\[
\alpha |s-s'| <  \frac{1}{B},
\]
are nearly parallel, with a small, rapidly fluctuating angle between
them. But rows that are far apart, with indices satisfying
\[
\alpha |s-s'| \geq \frac{3 \sqrt{2}}{ B},
\]
are essentially orthogonal, because the right hand side in
(\ref{eq:ANGLE}) is approximately zero. When the target is stationary,
and its cross-range offset is zero, then $\alpha = 0$, and all the
entries in $C$ are constant and equal to one. Moreover, all the rows
of $M$ are parallel to each other, and the matrix has rank one.  When
we view the data traces we see a vertical line at $t = 0$, as in the
left plot of Figure \ref{fig:traces}.  Obviously, the larger
$|\alpha|$ is, the closer the indices of the rows that are nearly
orthogonal. So we expect the rank of $M$ to increase with $|\alpha|$.
This is indeed the case, as we show next. We also illustrate this fact
in the right plot of Figure \ref{fig:traces}, where we show the matrix
of data traces for a stationary target that is offset from the
reference point $\vrho_o$.  Note that by definition (\ref{eq:Toep2}),
$|\alpha|$ is large when the cross-range offset of the target and/or
the target speed are large.

\subsubsection{Asymptotic characterization of the rank}
Because matrix $C$ is Toeplitz and large, we can use the asymptotic
Szeg\H{o} theory \cite{szego,bottcher} to estimate its rank.
To do so, let us define the sequence $\{c_j\}_{j\in \mathbb{Z}}$ with entries
\begin{equation}
  c_j =\frac{\sqrt{\pi}}{2B\Delta t}e^{-\frac{(\xi j)^2}{4}}\cos(\xi j), 
  \qquad \xi = B |\alpha| \Delta s.
\label{eq:defcj}
\end{equation}
A finite set of this sequence, for indices $|j| \le n$, defines
approximately the diagonals of the matrix $C$,
\begin{equation}
C_{j,l} \approx c_{j-l}, \quad j, l = 1, \ldots, n+1.
\end{equation}

Since multiplication of a large Toeplitz matrix with a vector is
approximately a convolution, and since convolutions are diagonalized
by the Fourier transform, it is not surprising that the spectrum of
$C$ is defined in terms of the symbol $Q(\theta)$, the coefficients of
the Fourier series of (\ref{eq:defcj}),
\begin{equation}
  Q(\theta)= \sum_{j=-\infty}^{\infty} c_j e^{ij\theta}, 
  \quad\quad \theta\in(-\pi,\pi).
\label{eq:series}
\end{equation}
With this symbol, we can characterize asymptotically in the limit $n
\to \infty$ the rank of $C$, using Szeg\H{o}'s first limit theorem
\cite{bottcher}, that gives  
\begin{equation}
\label{eq:RANK_AS}
  \lim_{n\to\infty} \frac{\cN(n;\beta_1,\beta_2)}{n+1}=
  \frac{1}{2\pi}\int_{-\pi}^{\pi}1_{[\beta_1,\beta_2]}(Q(\theta)) d\theta.
\end{equation}
Here $1_{[\beta_1,\beta_2]}$ is the indicator function of the interval
$[\beta_1,\beta_2]$ and $\cN(n;\beta_1,\beta_2)$ is the number of
eigenvalues of $C$ that lie in this interval.

We show in appendix \ref{sect:appSymbol} that the symbol is given
approximately by 
\begin{equation}
  Q(\theta)\approx \frac{\pi}{2B\Delta t\xi}\left\{\exp\left[-
      \frac{(\theta-\gamma)^2}{\xi^2}\right]+
    \exp\left[-\frac{(\theta+\gamma)^2}{\xi^2}\right]\right\},
\end{equation}
where $\gamma \in (-\pi,\pi)$ is defined by
\begin{equation}
  \gamma =[(\omega_0\alpha\Delta s+\pi)\mod 2\pi] - \pi.
\end{equation}
We use this result and (\ref{eq:RANK_AS}) to obtain an asymptotic
estimate of the essential rank, defined by 
\begin{equation}
  \mbox{rank}\left[C\right] := \cN\left(n;\epsilon
  \|Q\|_{\infty},\infty\right), \quad\quad \|Q\|_{\infty} = \sup_{\theta \in
    (-\pi,\pi)} |Q(\theta)|.
\end{equation}
Here $0< \epsilon \ll 1$ is a small threshold parameter, and
$\|Q\|_{\infty}$ is of the order of the largest singular value of
$C$. It follows from the Szeg\H{o} theory \cite{szego,bottcher} that
this singular value is given by the maximum of the symbol, which is of
the order of $\pi/(2 B \Delta t \xi)$.  We obtain that for $n \gg 1$,
\begin{eqnarray}
  \frac{\mbox{rank}\left[C\right]}{n+1}&\approx&
  \frac{1}{2\pi}\int_{-\pi}^{\pi} 1_{[ \epsilon \|Q\|_{\infty},
      \infty)} (Q(\theta)) d\theta \nonumber \\ &=&
    \min\left(\frac{2|\alpha| B\Delta s\sqrt{\log
        1/\epsilon}}{\pi},1\right),
\label{eq:AS_RANK}
\end{eqnarray}
where the last equality follows from direct calculation.

\begin{figure}[t]
\centering
\subfigure[Stationary]{\includegraphics[width=.45\columnwidth]{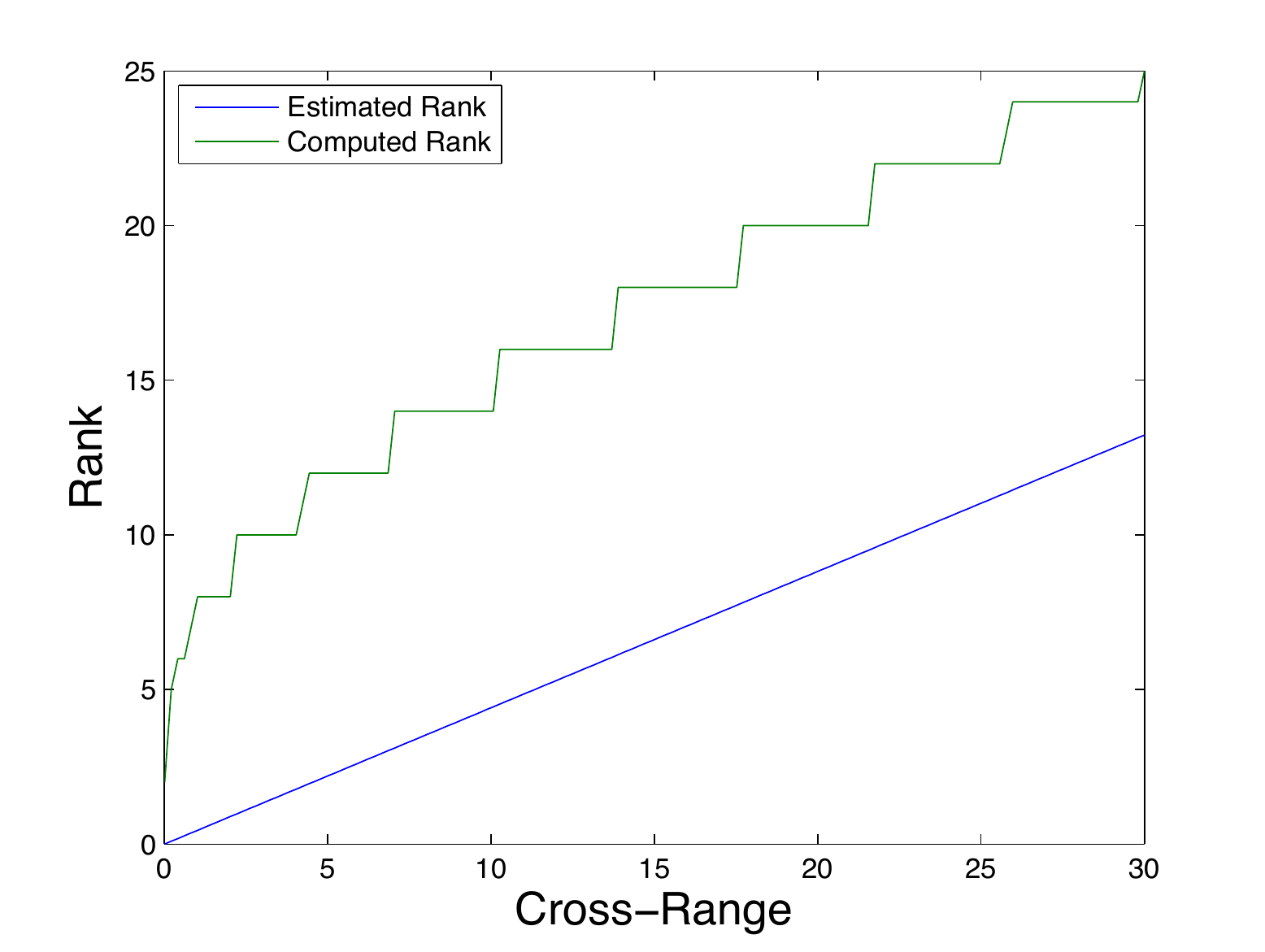}}
\subfigure[Moving]{\includegraphics[width=.45\columnwidth]{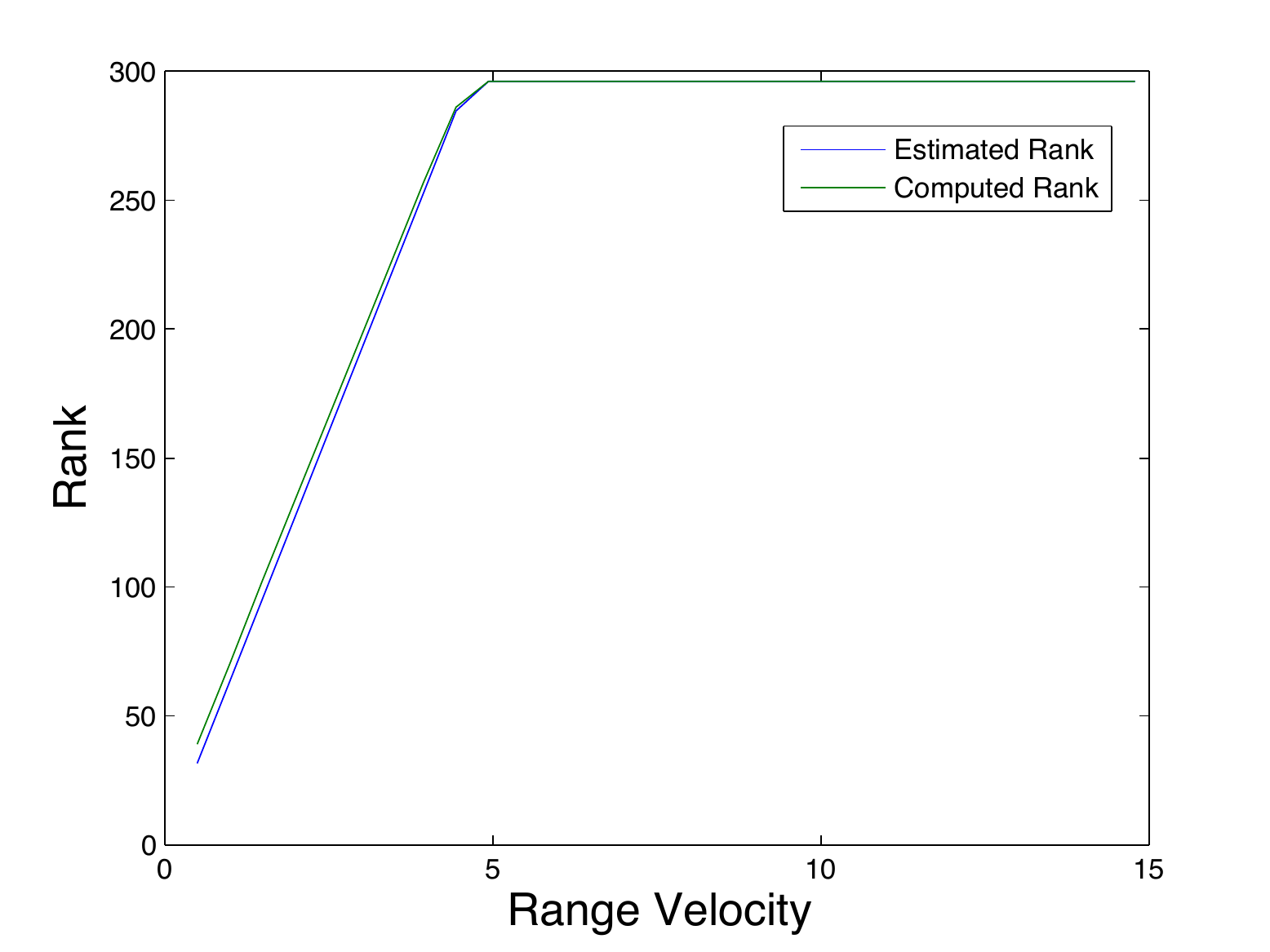}}
\caption{Comparison of the computed and estimated rank of $C$ for a
single target. Left: Stationary target at various cross-range positions.
Right: Moving target with various
velocities.}
\label{fig:rank1}
\end{figure}

\begin{figure}
\centering
\includegraphics[width=.5\columnwidth]{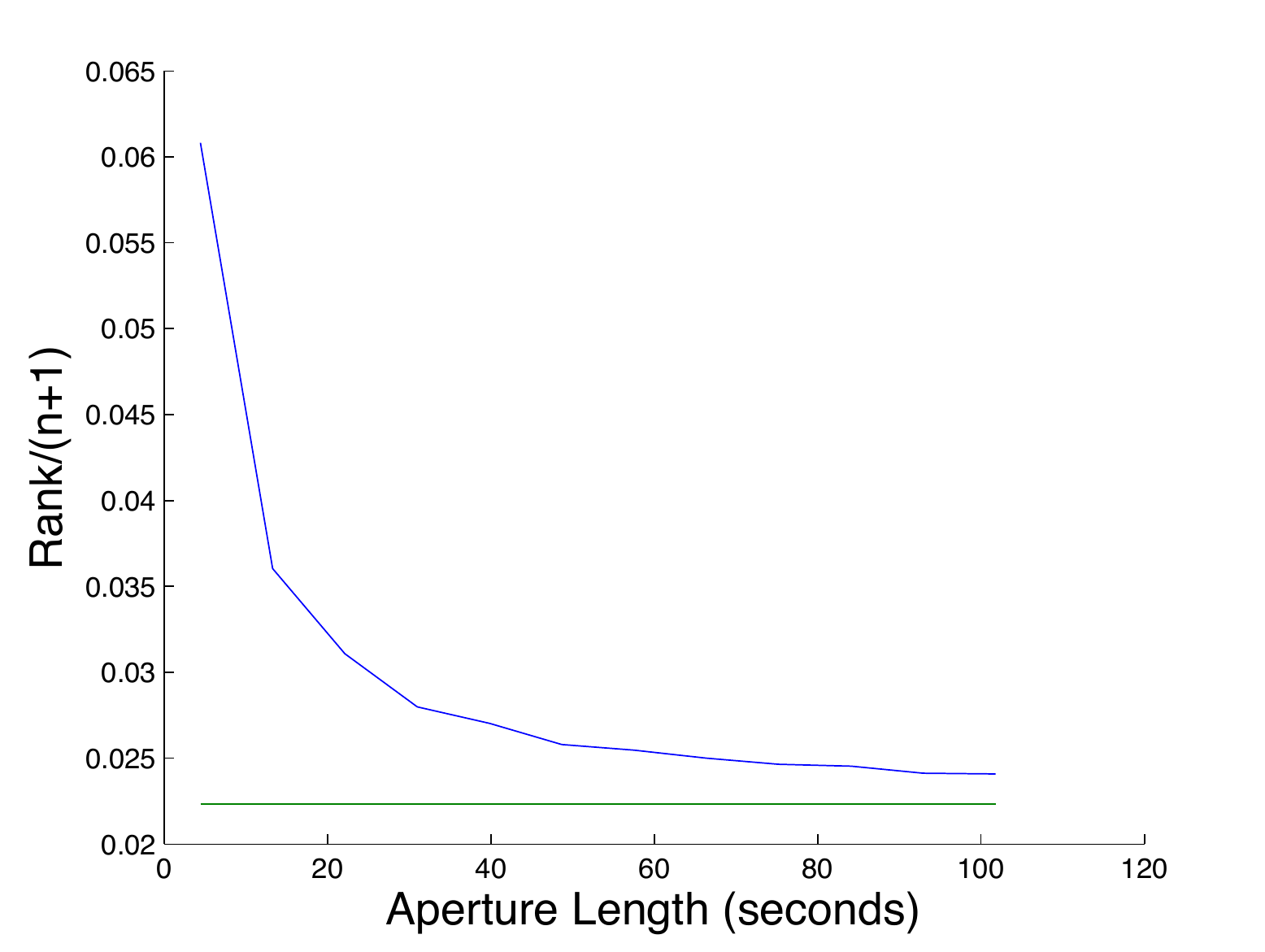}
\caption{Convergence of the rank of $C$ normalized by the size
  $(n+1)$. The blue line is the computed value and the green line is
  the asymptotic estimate.}
\label{fig:convergeRank}
\end{figure}

As an illustration, we show with green in Figure \ref{fig:rank1}a the
computed rank of the matrix $C$ for a single stationary target, versus
the cross-range position. Positions with larger cross-range result in 
larger $|\alpha|$ and thus larger rank, as expected. The asymptotic estimate
(\ref{eq:AS_RANK}) of the rank is shown in blue. It exhibits the same
linear growth in cross-range and therefore in $|\alpha|$, but it is
lower than the computed rank. This is because in this simulation $n$
is not sufficiently large. We get a good approximation when $c_j
\approx 0$ for $j \approx n$, so we can approximate the series
(\ref{eq:series}) that defines the symbol by the truncated sum for
indices $|j| \le n$. This is not the case in this simulation, so there
is a discrepancy in the estimated rank. However, the result improves
when we increase $n$, by increasing the aperture. This is illustrated
in Figure \ref{fig:convergeRank}, where we show the rank normalized by
the size of the matrix, and note that it has the predicted asymptote
as $n$ increases.

The plot in Figure \ref{fig:rank1}b shows the computed rank (in green)
and the asymptotic estimate (\ref{eq:AS_RANK}) (in blue) for a moving
target located at $\vrho = \vrho_o$ at $ s= 0$. We plot the rank as a
function of the range component of the velocity. The rank increases
with the velocity, and therefore with $|\alpha|$, as
expected. Moreover, the asymptotic estimate is very close to the
computed one, because in this case the entries in the sequence
$\{c_j\}_{j \in \mathbb{Z}}$ decay faster with $j$.  Finally, notice that even
for small velocities, the rank is much larger for the moving target than the 
stationary target.

\subsection{Analysis of the rank of the data traces for two targets}
\label{sect:double}
In the case of two targets at locations $\vrho_j(s)$ for $j = 1, 2$,
the entries of the matrix $M$ follow from equations (\ref{eq:MOD5})
and (\ref{eq:MOD6}),
\begin{equation}
  M_{jl} = \sum_{j=1}^2\cos\left[\om_o(t - \Delta\tau(s,\vrho_j(s)))\right]
  \exp \left[ -\frac{B^2}{2} (t - \Delta\tau(s,\vrho_j(s)))^2 \right].
\label{eq:TWO1}
\end{equation}
We take for simplicity the case of two stationary targets
\[
\vrho_j(s) = \vrho_j := \vrho_j(0), \quad j = 1, 2.
\] 
Extensions to moving targets with speeds $\vec{\bf u}_1$ and $\vec{\bf
u}_2$ are straightforward. They amount to redefining the parameters
$\alpha_1$ and $\alpha_2$ defined below by adding two linear terms in
the target velocity, as in equation (\ref{eq:Toep2}).

The expression of matrix $C = M M^T$ is given in the following 
proposition. It is obtained with a calculation that
is similar to that in appendix \ref{sect:appSingle}, using the same 
assumption on the Fresnel number as in Proposition \ref{prop:Toep}.

\begin{proposition}
Assume that the Fresnel number satisfies the bound (\ref{eq:Fresnel}),
with $\vu = 0$ since the targets are stationary.  The matrix $C$ has
entries defined by the function
\begin{align}
\cC(s,s')&\approx\frac{\sqrt{\pi}}{2B\Delta t}\left\{ \sum_{j=1}^2
\cos[\omega_o\alpha_j(s-s')]\exp \left[-\frac{(B\alpha_j)^2
(s-s')^2}{4}\right] \right.\nonumber\\ &+ \cos[\omega_o(\alpha_1
s-\alpha_2 s'+\beta)] \exp\left[-\frac{B^2(\alpha_1 s-\alpha_2
s'+\beta)^2}{4}\right]\label{eq:covar2}\\ &\left. +
\cos[\omega_o(\alpha_1 s'-\alpha_2 s+\beta )]
\exp\left[-\frac{B^2(\alpha_1 s'-\alpha_{2}
s+\beta)^2}{4}\right]\right\},\nonumber
\end{align}
sampled at the discrete slow times. Here we let
\begin{equation}
\alpha_j =-\frac{2V \vt \cdot \mathbb{P}_o
(\vrho_j-\vrho_o)}{cL},
\label{eq:alphaj}
\end{equation}
and 
\begin{equation}
\beta = \frac{2}{c}\sum_{j=1}^2 (-1)^{j} \left\{ \vm_o \cdot
(\vrho_j-\vrho_o) + \frac{[\vm_o \cdot (\vrho_j-\vrho_o)]^2}{2 L}
\right\}.
\label{eq:beta}
\end{equation}
\end{proposition}

The structure of the matrix $C$ is now more complicated. The first
term in (\ref{eq:covar2}) gives a Toeplitz matrix, as before. The
other two, which are due to the interaction between the targets, give
matrices that are approximately g-Toeplitz or g-Hankel, depending on
the sign of the ratio $\alpha_2/\alpha_1$.

\begin{definition}
An $(n+1) \times (n+1)$ g-Hankel matrix  $H$ with shift $g \in
\mathbb{Z}^{+}$ is defined by a sequence $\{h_j\}_{j \in \mathbb{N}}$ as 
\[
H_{jl} = h_{j-1+g(l-1)}, \qquad j, l = 1,\ldots, n+1.
\]
The matrix is Hankel when $g = 1$.  A g-Toeplitz matrix is defined
similarly, by replacing $g$ with $-g$.
\end{definition}

\begin{figure}[t]
\hspace{-0.9in}\includegraphics[width=1.27\columnwidth]{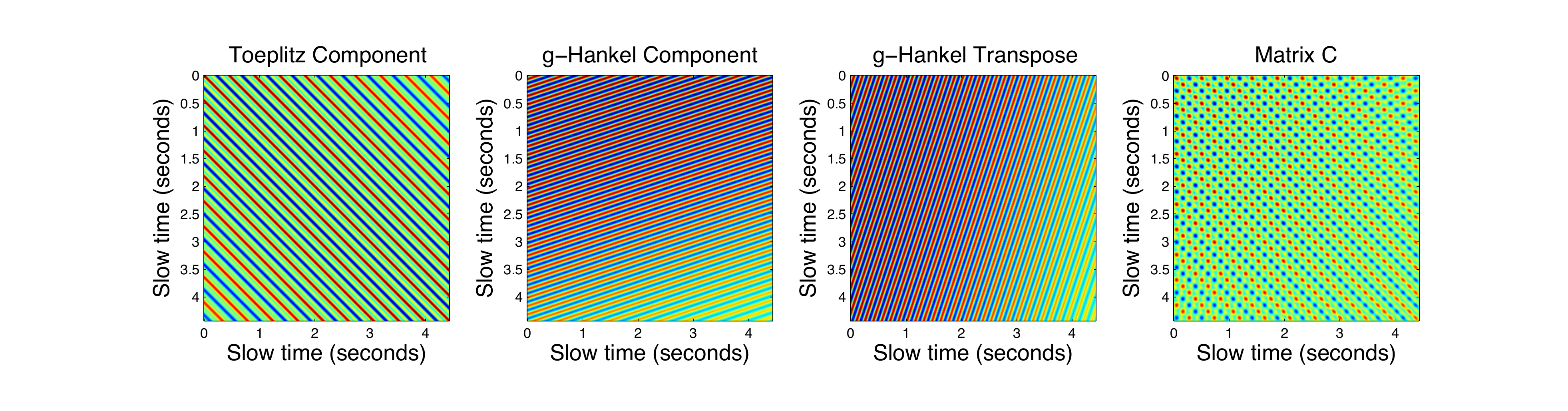}
\caption{Components of matrix $C$ for two stationary targets located
at $\vrho_1 = (0.15,15,0)$m and $\vrho_2=(-0.15,-5,0)$m. On the left we
show the Toeplitz part $T$. The next two plots show the g-Hankel part
$H$ and its transpose $H^T$. The right plot shows the sum, i.e., the
matrix $C$.}
\label{fig:toeplitzHankel}
\end{figure}

To analyze the spectrum of matrix $C$, we choose the reference 
point $\vrho_o$ in such a way that 
\begin{equation}
\frac{\alpha_2}{\alpha_1} < 0, \quad \mbox{and}
\quad g:=\left|\frac{\alpha_2}{\alpha_1}\right| \in \mathbb{N}.
\label{eq:assumerho_o}
\end{equation}
Then, $C$ is given by 
\begin{equation}
C = T + H + H^T,\label{eq:sumTH}
\end{equation}
with Toeplitz matrix
\begin{equation}
\label{eq:Tg}
T_{jl} = c_{j-l}, \qquad j, l = 1, \ldots, n+1,
\end{equation}
and g-Hankel matrix
\begin{equation}
H_{jl} = h_{(j-1)+g(l-1)}, \qquad j, l = 1, \ldots, n+1.
\label{eq:Hg}
\end{equation}
Here $\{c_j\}_{j \in \mathbb{Z}}$ and $\{h_j\}_{j \in \mathbb{N}}$ 
are sequences with entries
\begin{equation}
c_j = \frac{\sqrt{\pi}}{2B\Delta t}\left[e^{-\frac{(\xi_1
  j)^2}{4}}\cos(\xi_1 j) +e^{-\frac{(\xi_2 j)^2}{4}}\cos(\xi_2
  j)\right] , \quad \xi_{\ell} = B |\alpha_{\ell}| \Delta s, \quad \ell=1,2
\label{eq:seqc}
\end{equation}
and 
\begin{equation}
h_j = \frac{\sqrt{\pi}}{2B\Delta t}e^{-\frac{[\xi_1 (j +
  \zeta)]^2}{4}}\cos[\xi_1 (j + \zeta)], \qquad \zeta =
  \frac{\beta}{|\alpha_1| \Delta s}.
\label{eq:seqh}
\end{equation}
An example of matrix $C$ and its Toeplitz and g-Hankel parts is 
shown in Figure \ref{fig:toeplitzHankel}.

If the range offsets are large, so that
\[
\xi_1 |\zeta| = B |\beta| = \frac{2 B}{c} 
\left|\sum_{j=1}^2 (-1)^j \{ \vm_o \cdot (\vrho_j - \vrho_o) + 
\frac{[\vm_o \cdot (\vrho_j - \vrho_o)]^2}{2 L} \right| \gg 1,
\]
the g-Hankel matrix has small entries, and $C$ is approximately
Toeplitz, as in the single target case. The difference of the travel
times between the SAR platform and such targets is larger than the
compressed pulse width, and their interaction in (\ref{eq:covar2}) is
negligible.  If the range offsets are small, the structure of the
matrix $C$ is as in equation (\ref{eq:sumTH}), and the estimate of its
rank follows from the recent results in
\cite{sesana,fasino,tilli}. They say that the g-Hankel terms $H + H^T$
have a negligible effect on the rank in the limit $n \to \infty$.
See appendix \ref{sect:gHankel} for more details.
Thus, in either case, the rank estimate of $C$ is given by equation
(\ref{eq:RANK_AS}), in terms of the symbol $Q(\theta)$ defined by
(\ref{eq:series}), using the sequence $\{c_j\}_{j\in \mathbb{Z}}$ with
entries (\ref{eq:seqc}). 

Explicitly, the symbol is given by 
\begin{equation}
\label{eq:2TSYMB}
Q(\theta)\approx \frac{\pi}{2B\Delta t\xi_1}\left[e^{-
    \frac{(\theta-\gamma_1)^2}{\xi_1^2}}+
  e^{-\frac{(\theta+\gamma_1)^2}{\xi_1^2}}\right] +
\frac{\pi}{2B\Delta t\xi_2}\left[e^{-
    \frac{(\theta-\gamma_2)^2}{\xi_2^2}}+
  e^{-\frac{(\theta+\gamma_2)^2}{\xi_2^2}}\right],
\end{equation}
with $\gamma_j \in (-\pi,\pi)$ defined by
\begin{equation}
  \gamma_j =[(\omega_0\alpha_j\Delta s+\pi)\mod 2\pi] - \pi,
\end{equation}
for $j = 1,2$, and the rank is given by
\begin{eqnarray}
  \mbox{rank}\left[C\right] \approx
  \frac{(n+1)}{2\pi}\int_{-\pi}^{\pi} 1_{[ \epsilon \|Q\|_{\infty},\infty)} (Q(\theta)) d\theta.
\end{eqnarray}

\begin{figure}[t]
\centering
\includegraphics[width=.9\columnwidth]{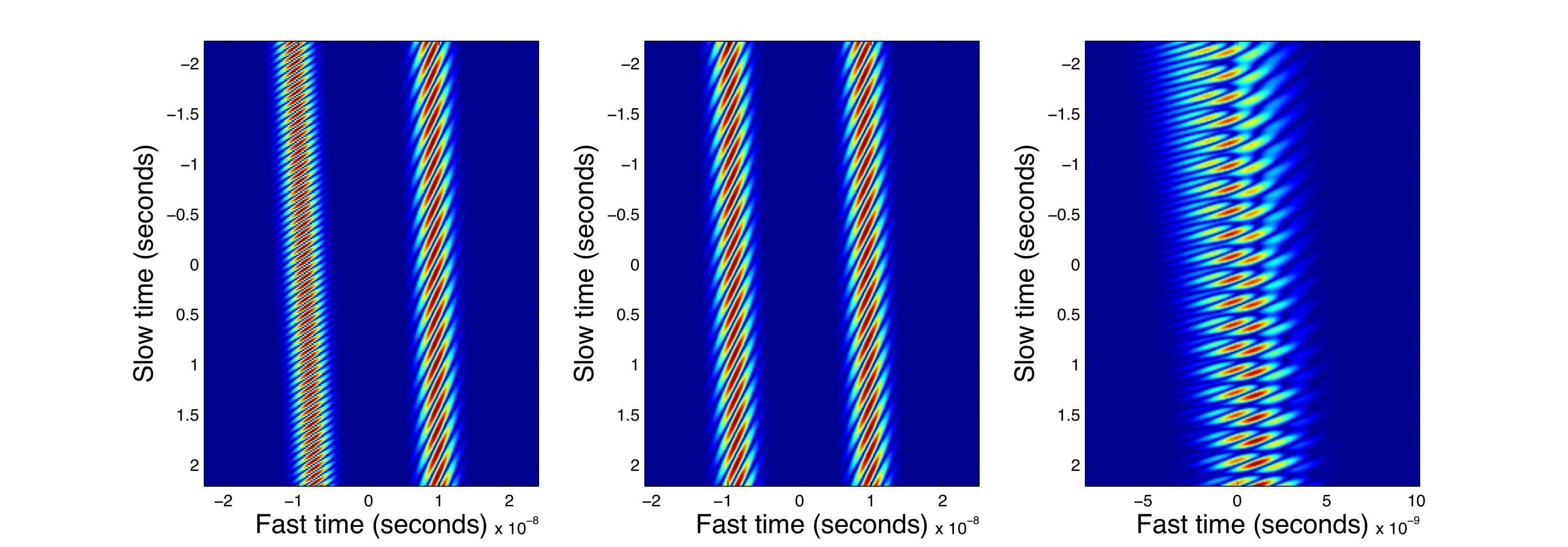}
\caption{Data traces from two stationary targets. Left: the targets
  are at positions $\vrho_1 = (2,5,0)$m and $\vrho_2 = (-2,15,0)$m.
  Middle:  $\vrho_1 = (2,5,0)$m and
  $\vrho_2 = (-2,5,0)$m.  Right: $\vrho_1 = (0.15,5,0)$m and $\vrho_2
  = (-0.15,15,0)$m.}
\label{fig:traces2}
\end{figure}

\subsubsection{Illustration} 

\begin{figure}[t]
\centering
\includegraphics[width=.9\columnwidth]{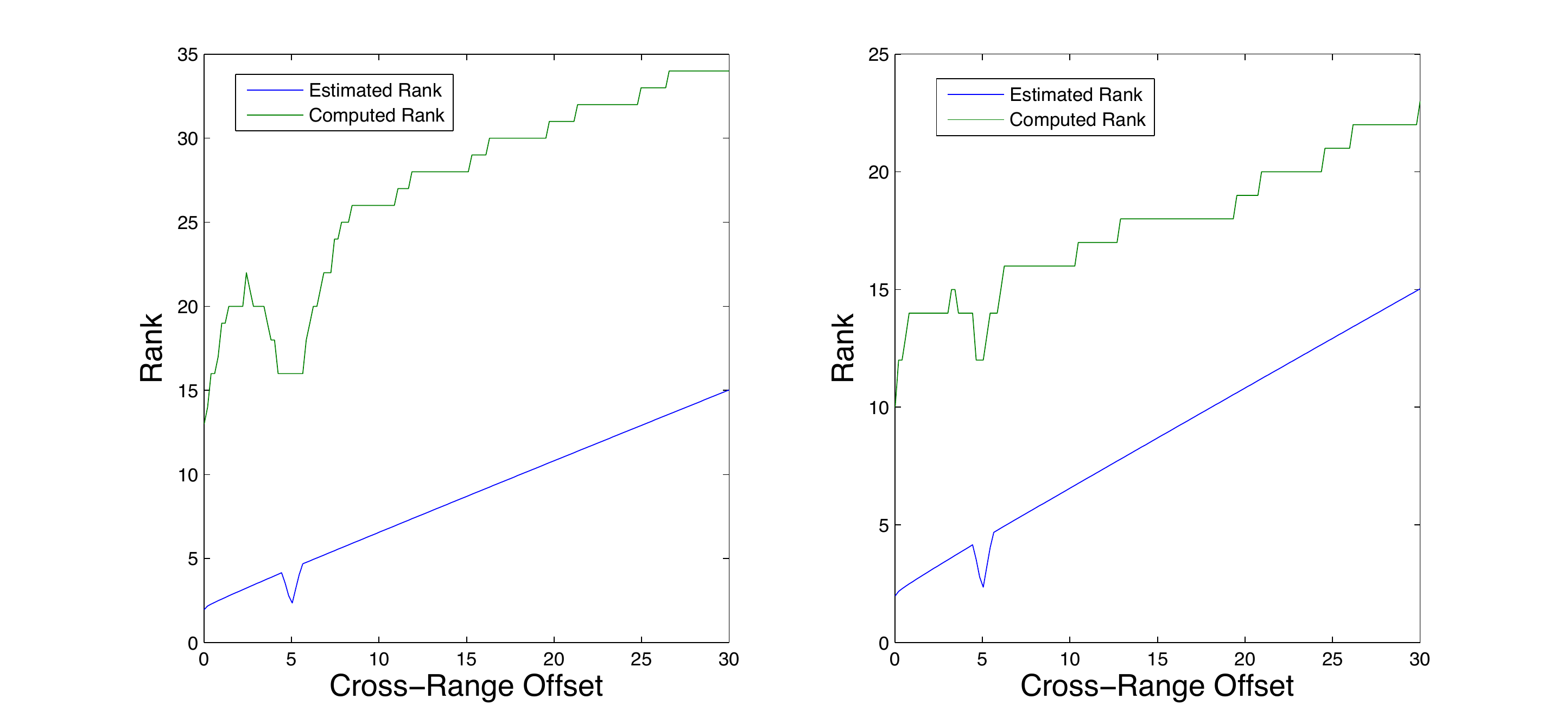}
\caption{Computed and estimated rank of matrix $C$ for two stationary
targets. One is fixed at location $\vrho_1 =(5,5,0)$m and the location
of the other varies on the line segment between $(-5,0.01,0)$m and 
$(-5,30,0)$m.}
\label{fig:eigenvals2}
\end{figure}

We compare in Figure \ref{fig:eigenvals2} the computed and estimated
rank for two stationary targets. The results on the left are for one
target fixed at location $\vrho_1 = (5,5,0)$m. We vary the location of
the other target between $(-5,0.01,0)$m and $(-5,30,0)$m. The range
separation is $10$m, so that $ B |\beta| = 41.47, $ and the g-Hankel
matrix $H$ has negligible entries. The results on the right are for
one target at $\vrho_1 = (0.15,5,0)$m and the location of the other
varying between $(-0.15,0.01,0)$m to $(-0.15,30,0)$m. The range
separation is $0.3$m, so that $ B |\beta| = 1.24, $ and the g-Hankel
matrix $H$ is no longer negligible. We see that in spite of $H$ being
neglible or not, the rank of matrix $C$ behaves essentially the same,
as predicted by the asymptotic theory. This would be difficult to
guess by just looking at the data traces displayed in Figure
\ref{fig:traces2}.  

The computed and the estimated ranks grow at the same rate with the
cross range of the second target, i.e., with $|\alpha_2|$.  The growth
is monotone except in the vicinity of the local minimum corresponding
to the targets having exactly the same cross-range. In this special
configuration $\alpha_1 = \alpha_2$, and therefore $\xi_1 = \xi_2$.
The symbol (\ref{eq:2TSYMB}) simplifies to 
\[
Q(\theta)\approx \frac{\pi}{B\Delta t\xi_1}\left[e^{-
    \frac{(\theta-\gamma_1)^2}{\xi_1^2}}+
  e^{-\frac{(\theta+\gamma_1)^2}{\xi_1^2}}\right],
\]
and it exceeds the threshold $\epsilon \|Q\|_{\infty}$ for $\theta$ in
a smaller subset of $(-\pi,\pi)$, than in the general case with $\xi_1
\ne \xi_2$. The rank is defined by the size of this set, so it should
have a minimum as observed in Figure \ref{fig:traces2}.

Similar to the result in Figure \ref{fig:rank1} for a single
stationary target, there is a discrepancy between the computed and
estimated rank, due to the aperture not being large enough. This
discrepancy diminishes as we increase $n$ and therefore the aperture.

\subsection{Discussion}
\label{sect:rankDisc}
The analysis above shows that the matrix of data traces from a moving
target has much higher rank than that from a stationary target.  The
larger the speed, the higher the rank. The rank of the matrix of
traces from a stationary target is smallest, equal to one, when the
target is at the same cross-range as the reference point $\vrho_o$.
The rank increases at a linear rate with the cross-range offset from
$\vrho_o$.

Comparing the results for one and two stationary targets, we see that
the rank increases. The rank depends strongly on the cross-range
offset of the targets. There is a small effect due to the separation
of the targets in range, but it becomes negligible in the asymptotic
limit $n \to \infty$.

Although we have not presented an analysis for more than two targets,
we observe numerically that the rank increases as we add more and more
stationary targets. The implication is that the matrix of traces from
a stationary scene with many targets is not in general low rank. This
is why the data separation with robust PCA should be done in
successive small time windows, with each window containing the traces
from only a few stationary targets. These traces give a matrix that is
low rank, and thus can be separated from the traces due to moving
targets. The simulation results shown in Figure \ref{fig:rpcaEx2b} 
illustrate this point.

\section{Numerical simulations}
\label{sect:numerics}
\setcounter{equation}{0}

We begin with the setup for the numerical simulations. Then we present
three sets of results.

\subsection{Setup}
\label{sect:setup}
We use the GOTCHA Volumetric SAR setup described in section
(\ref{sect:scaling}).  The data traces are generated with the model
(\ref{eq:MOD1}). In all the simulations but the last one, the point
targets are assumed identical, with reflectivity $\sigma_q = 1$.  The
images are obtained by computing the function (\ref{eq:imagefxn}) at
points in the square imaging region of area $70 \times 70 \, \mbox{m}^2$
centered at $\vrho_o$. The motion estimation results are obtained with
the phase space algorithm introduced in \cite{sar}. This algorithm
requires that we know the location of the target at one instant. We
choose it at the center of the aperture, which is why there is no
error in the target trajectory at $s = 0$.

The principle component pursuit optimization in the robust PCA is
solved with an augmented Lagrangian approach.  It requires the
computation of the top few singular values and corresponding singular 
vectors of large and sparse matrices, which we do with the software 
package PROPACK.

\subsection{Simulation 1}
\label{sect:sim1}
The first simulation is for a collection of $30$ stationary targets
placed randomly in the imaging region, and a single moving target
located at $(0,0,0)$m at $s = 0$, and moving in the plane with
velocity ${\bf u} = \frac{28}{\sqrt{2}}(1,1)$m/s.  The data traces are
shown in Figure \ref{fig:rpcaEx2a}.

\begin{figure}[!h]
\centering
\includegraphics[width=.9\columnwidth]{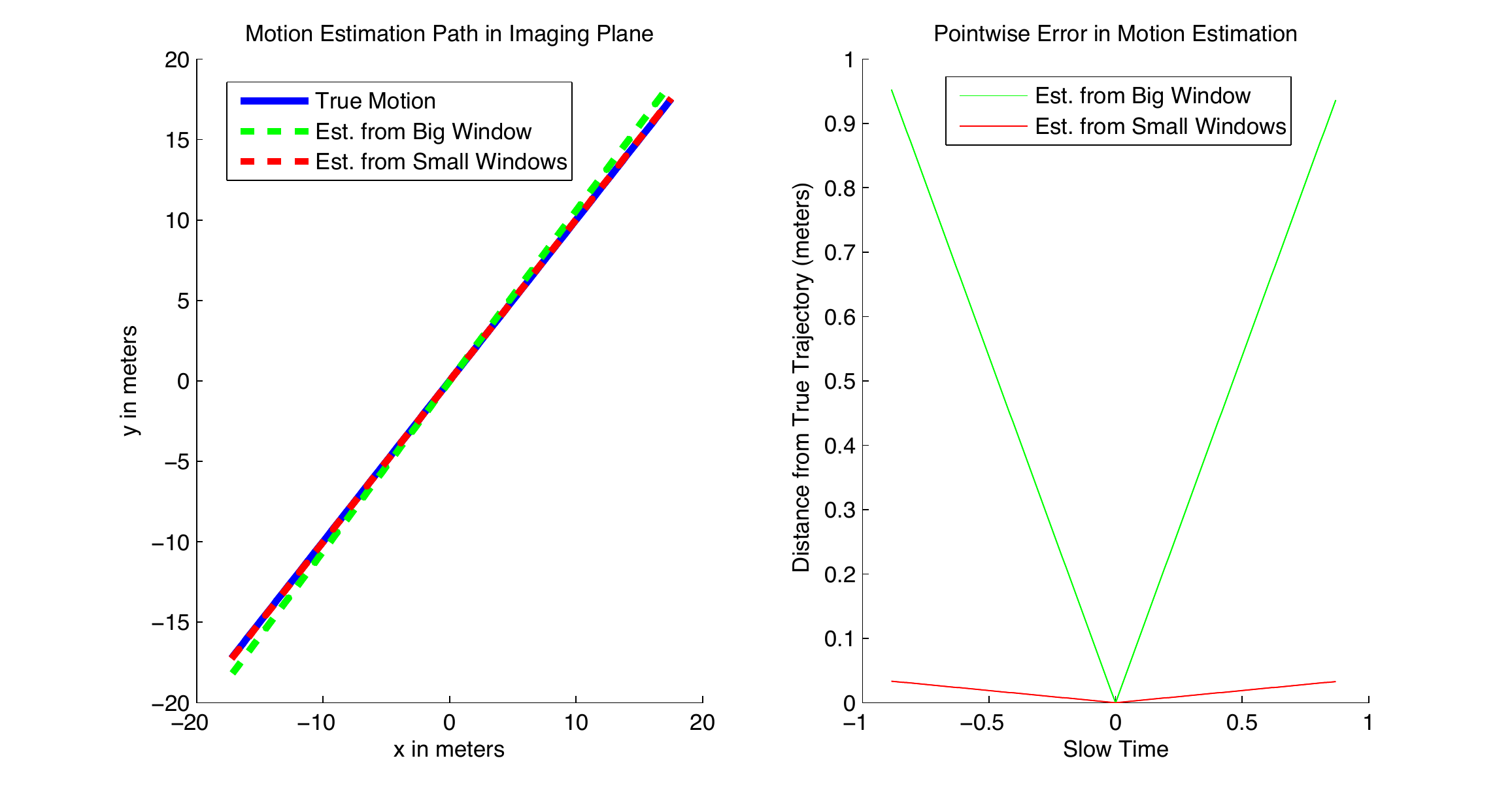}
\caption{Estimation of the trajectory of a moving target in a complex
  scene with 30 stationary scatterers, placed at random in a $50\times
  50\ m^2$ imaging region.  We compare the results obtained with the
  sparse part of the data traces returned by robust PCA with and
  without windowing.  These sparse parts are shown in the top right and
  bottom right plots in \ref{fig:rpcaEx2b}.}
\label{fig:motEst30}
\end{figure}

The estimated moving target trajectory is shown in the left plot in
Figure \ref{fig:motEst30}. The blue line corresponds to the true
trajectory.  The red and green lines are the estimated trajectories
with the sparse component of the matrix of traces, as returned by
robust PCA with and without windowing. These sparse components are
shown in the right plots in Figure \ref{fig:rpcaEx2b}. The separation
with robust PCA is better for the windowed traces, and so is the
estimate of the target trajectory. This is more clear in the rigÄht
plot of Figure \ref{fig:motEst30}, where we show the error of the
trajectory.

\begin{figure}[!h]
\hspace{-.1in}
\subfigure{\includegraphics[width=.53\columnwidth]{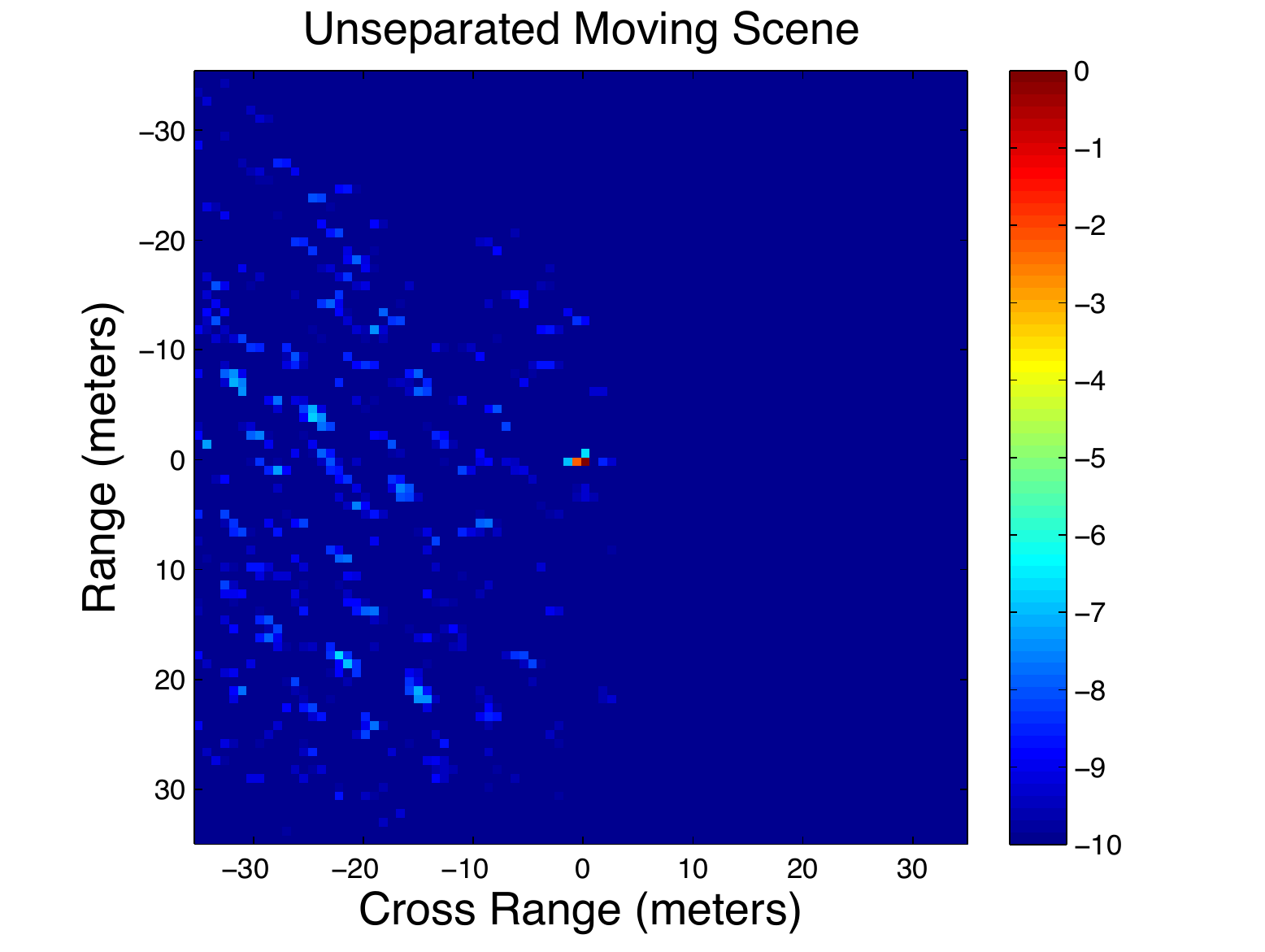}}\hspace{-.25in}
\subfigure{\includegraphics[width=.53\columnwidth]{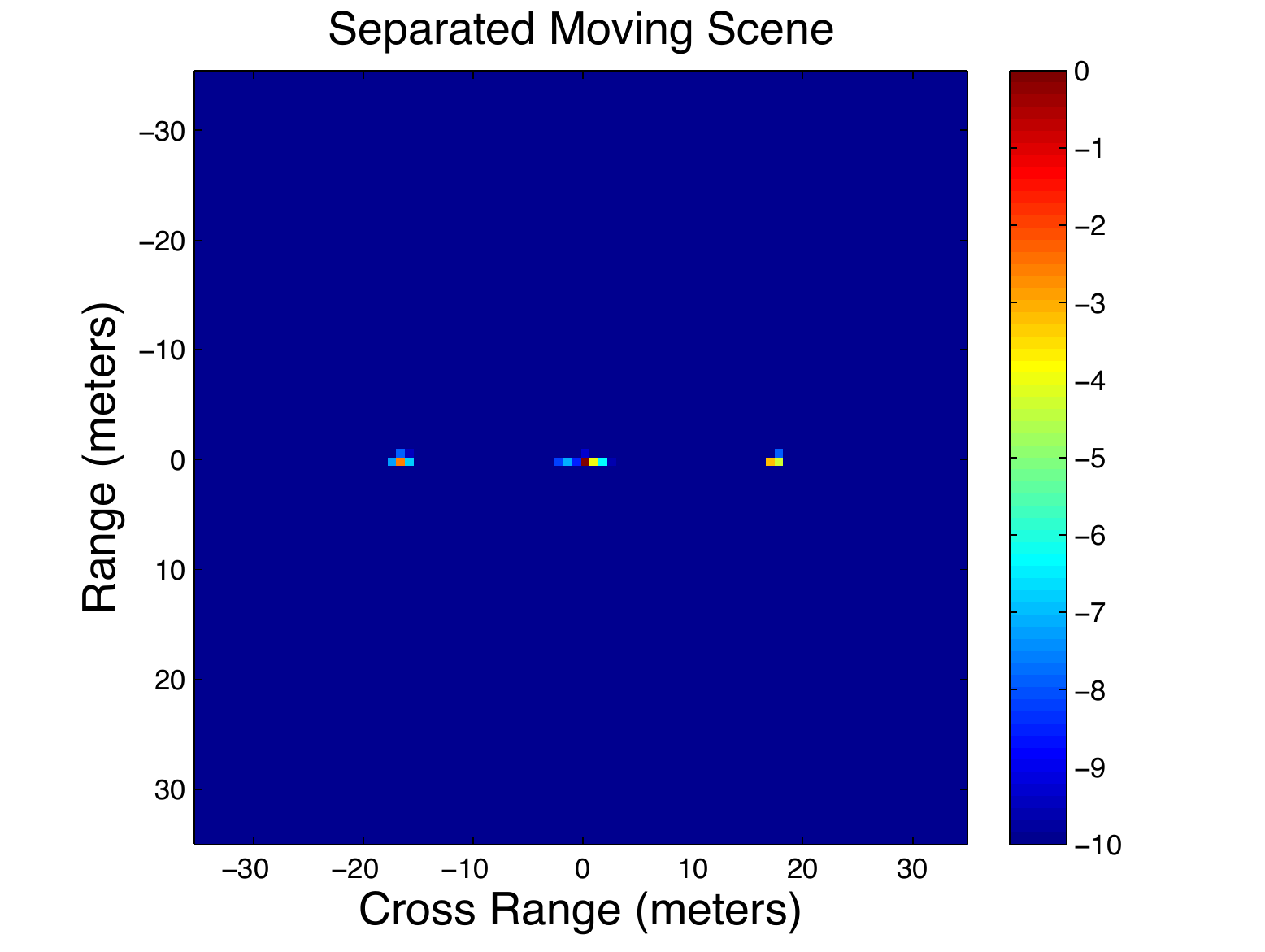}}\hspace{-.25in}
\caption{Images of a scene with 30 randomly placed stationary
  scatterers and one moving target with velocity 28 m/s.  Left: the
  image given by the original data with exact compensation of the
  motion of the target.  Right: the image given by the sparse
  component of the data, separated from the other traces by robust PCA
  with time windowing.  The images are normalized by the largest pixel value and plotted in dB.}
\label{fig:imagesRegMov}
\end{figure}

In the left plot of Figure \ref{fig:imagesRegMov} we show the image
obtained with the data traces, and with exact compensation of the
motion of the target. The image is focused at the initial location
$(0,0,0)$m of the moving target, as expected. However, the stationary
targets are out of focus, and the image appears noisy. The right plot
of Figure \ref{fig:imagesRegMov} shows the image obtained with the
sparse component of the traces, separated successfully by robust PCA
with data windowing. The motion compensation is with the estimated
velocity. We note that the artifacts due to the stationary targets are
now removed, and the image peaks at the expected location $(0,0,0)$m.
There are two ghost peaks, due to the error in the estimated target
velocity, but they are much smaller than the peak at the correct
location.

\subsection{Simulation 2}
\label{sect:sim2}
The second simulation is for a scene with 20 stationary targets and
two moving targets. The first moving target is as in the first
simulation, The second one is located at $(-5,5,0)$m at $s = 0$ and
moves in the plane with velocity ${\bf u} = \frac{14}{\sqrt{3}}
\left(-1,\sqrt{2}\right)$m/s.

\begin{figure}[!t]
\subfigure{\includegraphics[width=0.36\columnwidth]{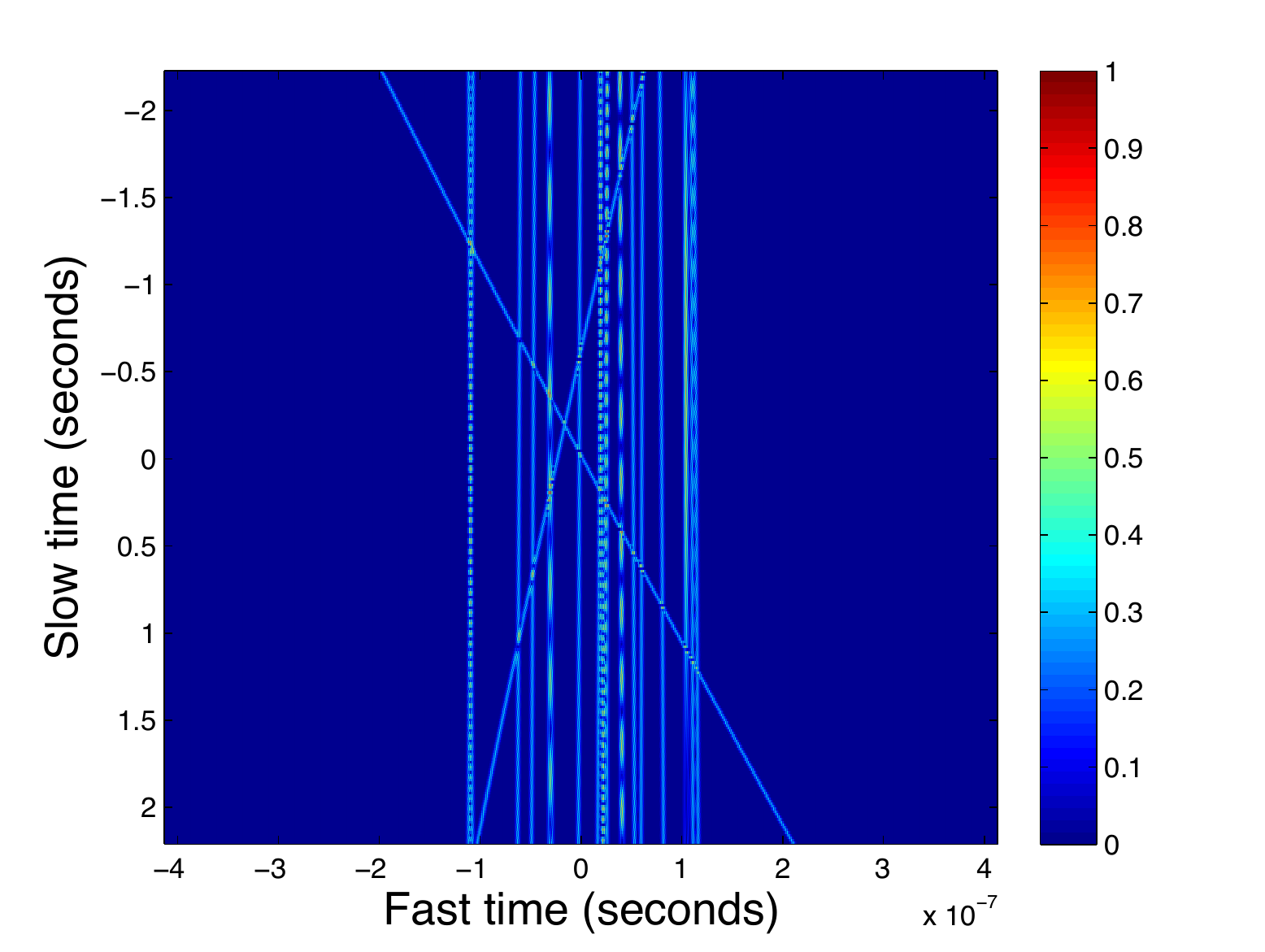}}\hspace{-.25in}
\subfigure{\includegraphics[width=0.36\columnwidth]{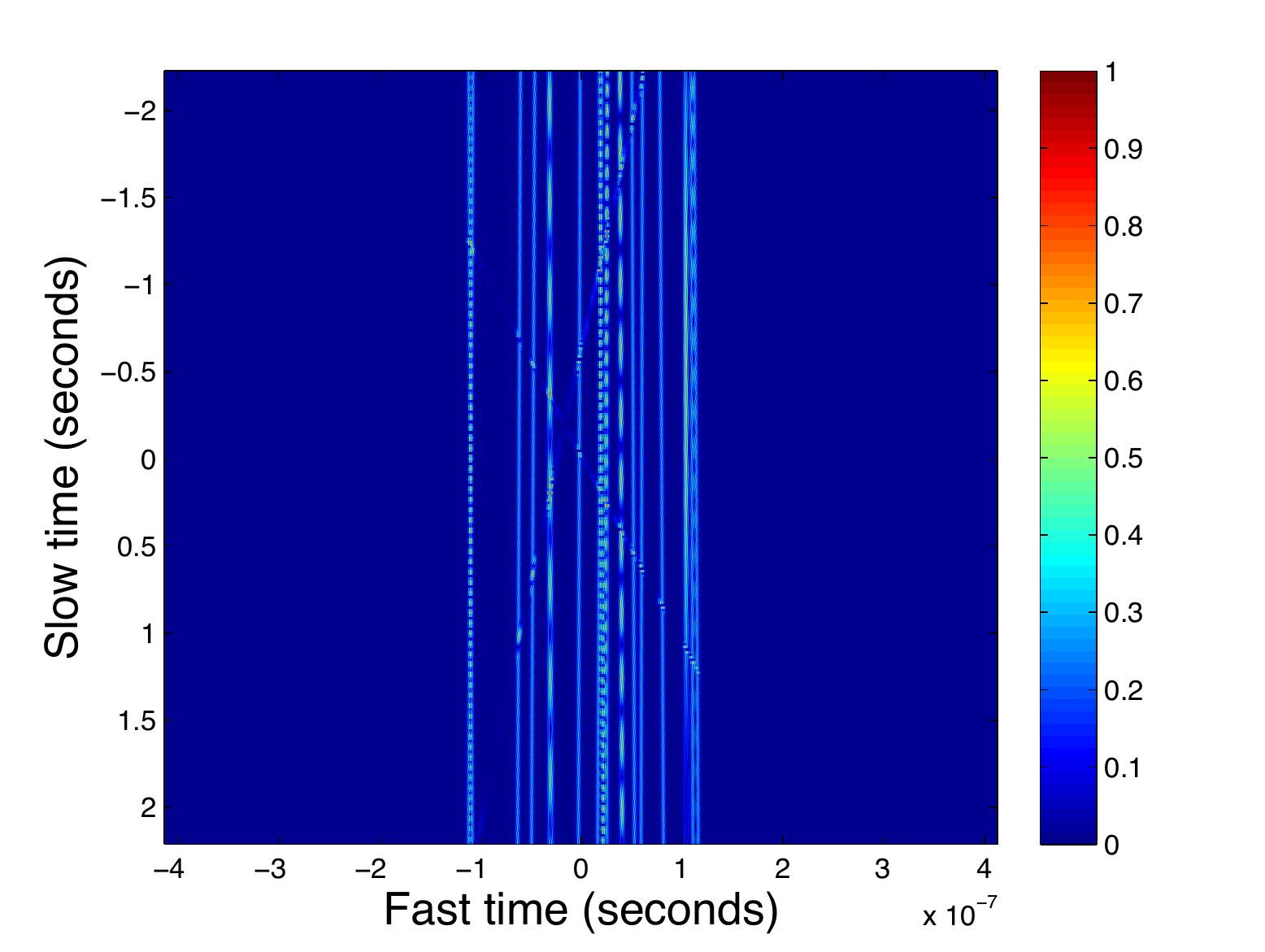}}\hspace{-.25in}
\subfigure{\includegraphics[width=0.36\columnwidth]{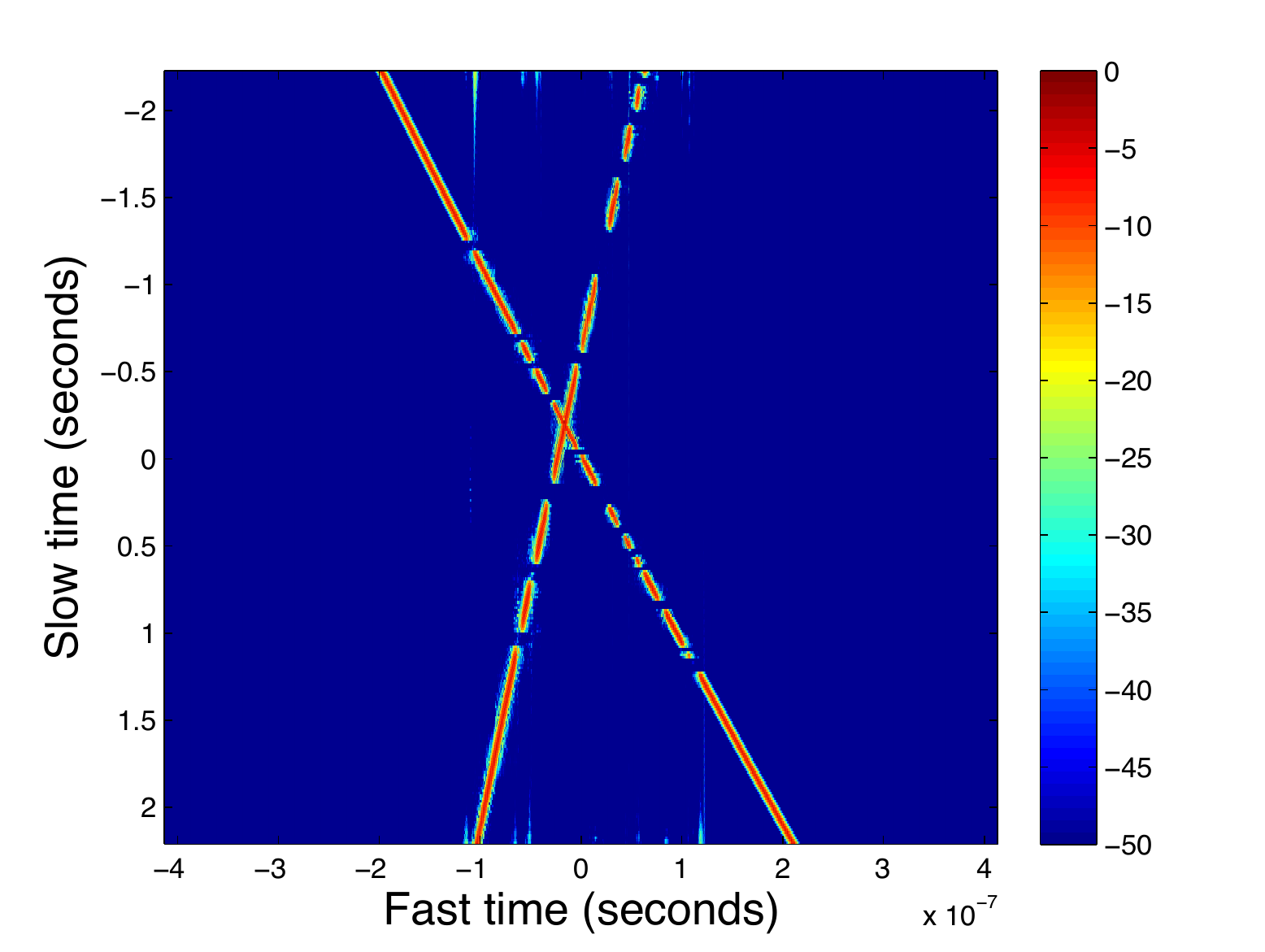}}
\caption{Data trace separation with robust PCA for a scene with 20
  stationary targets and two moving targets. From left to right we
  show the matrix $M$ of data traces, the low rank $\cL$, and the
  sparse $\cS$ parts. In each plot we show absolute values normalized
  by the largest value of $|D_r(s,t)|$.  The sparse component is
  plotted in dB scale to emphasize the contrast.}
\label{fig:rpca2mov}
\end{figure}

The data traces $D_r(s,t)$ are plotted on the left in Figure
\ref{fig:rpca2mov}.  The separation with robust PCA is shown in the
middle and right plots of Figure \ref{fig:rpca2mov}.  Each is
normalized by the maximum of $|D_r(s,t)|$, and then plotted on the
same color scale. Note how the traces from the two moving targets are
separated from the other traces. This simplifies the motion
estimation.

\subsection{Simulation 3}

Our last simulation considers again a scene with 30 stationary targets
and a single moving target.  This target is like that in simulation
one, except that its reflectivity is ten times larger than that of the
stationary targets.

\begin{figure}[!h]
\subfigure{\includegraphics[width=0.36\columnwidth]{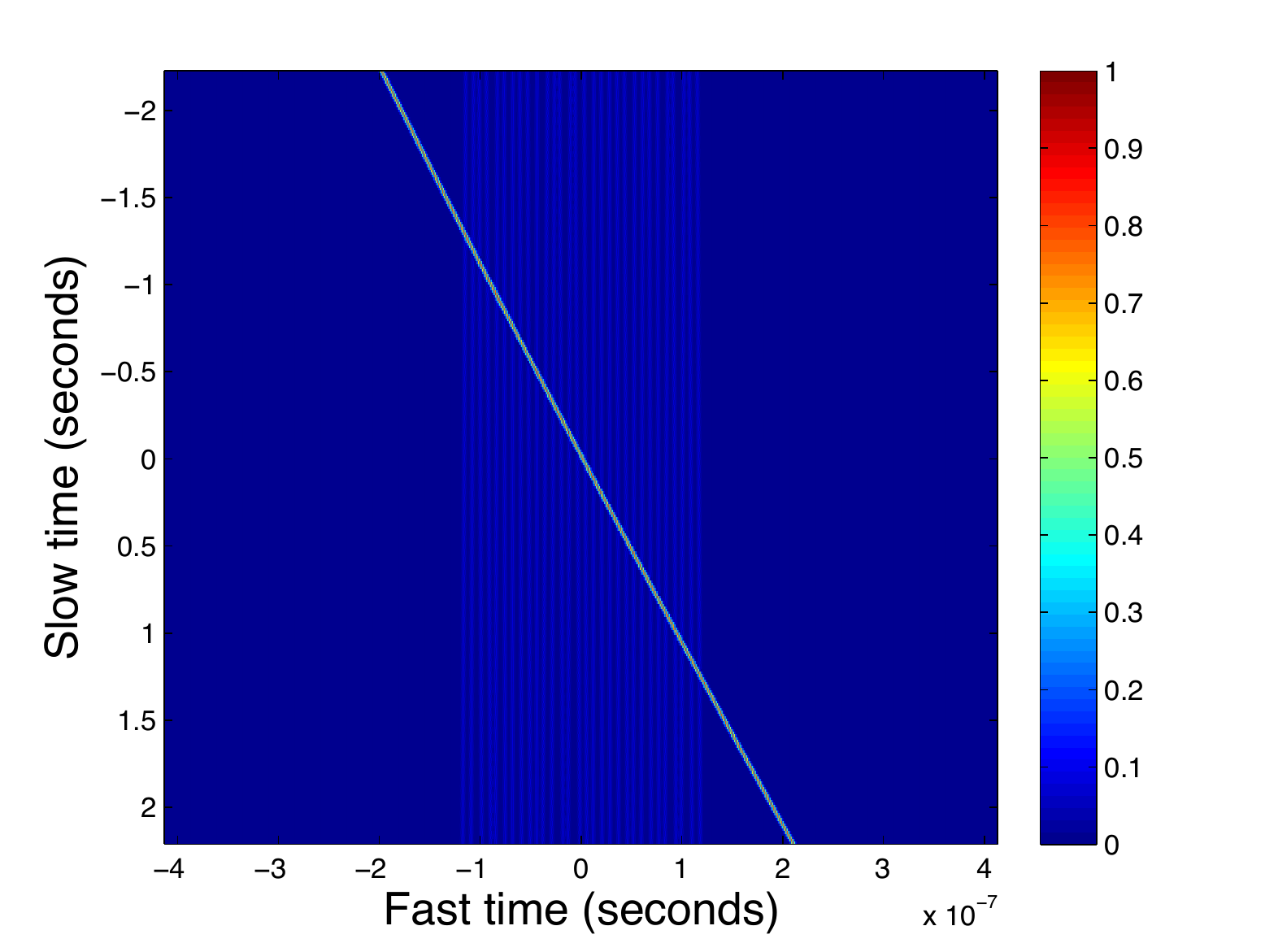}}\hspace{-.25in}
\subfigure{\includegraphics[width=0.36\columnwidth]{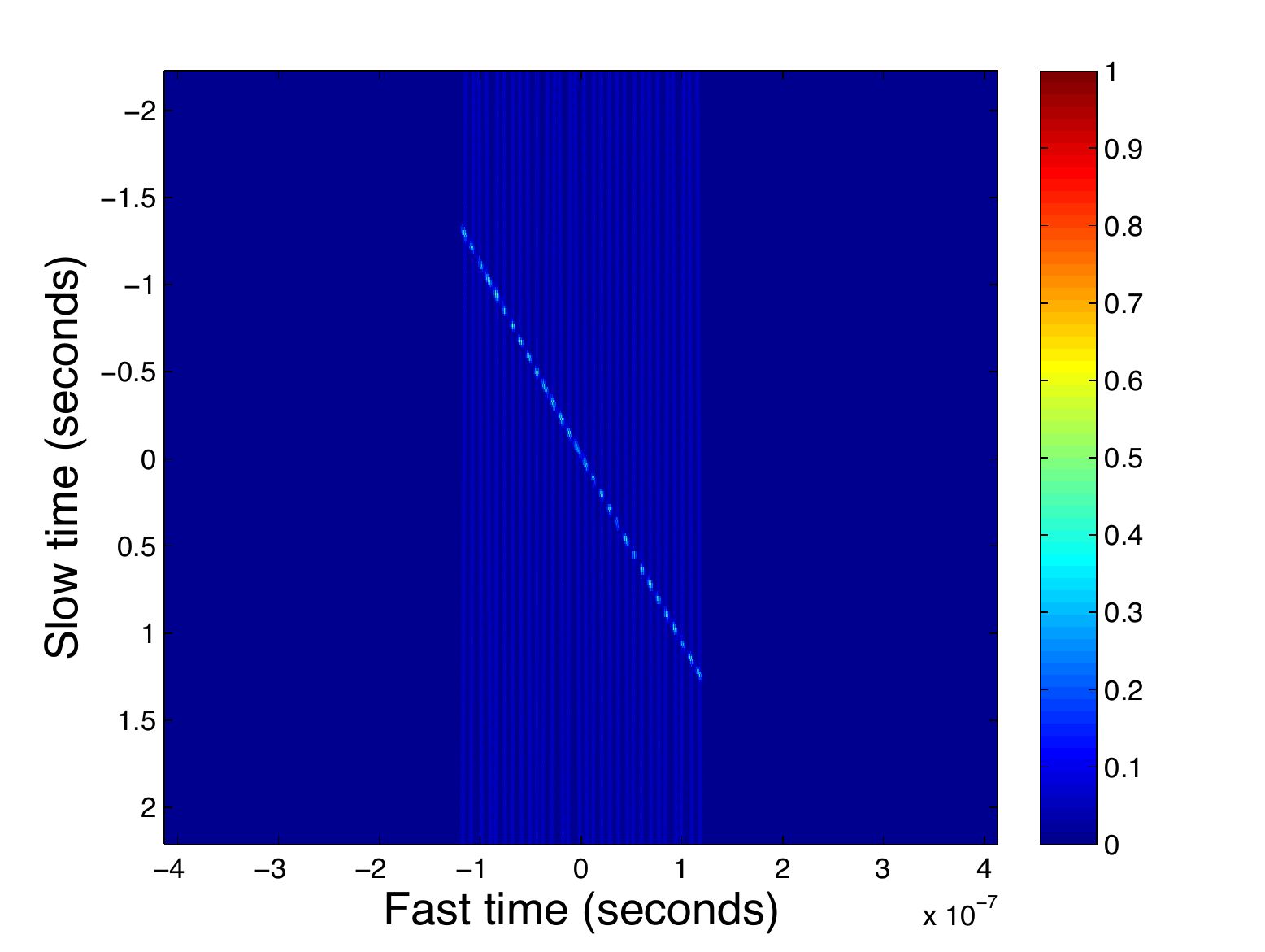}}\hspace{-.25in}
\subfigure{\includegraphics[width=0.36\columnwidth]{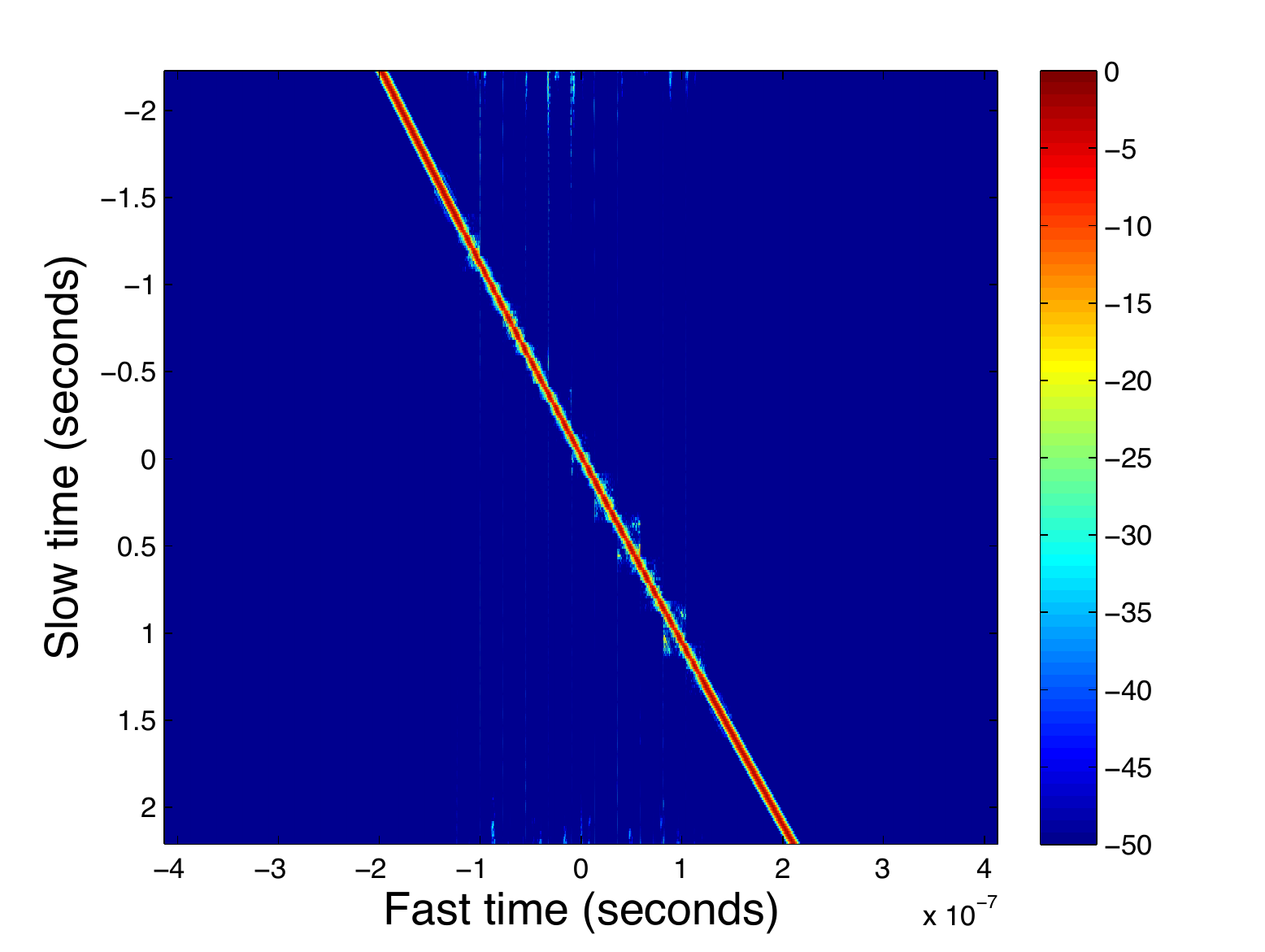}}
\caption{Data trace separation with robust PCA for a scene with 30
  stationary targets and a moving one with reflectivity that is ten
  times stronger than the others.  From left to right we show the
  matrix $M$ of data traces, the low rank $\cL$, and the sparse $\cS$
  parts. In each plot we show absolute values normalized by the
  largest value of $|D_r(s,t)|$.  The sparse component is plotted in
  dB scale to emphasize the contrast.  }
\label{fig:rpcaStrongMov}
\end{figure}

\begin{figure}[t]
\hspace{-.7in}
\includegraphics[width=1.2\columnwidth]{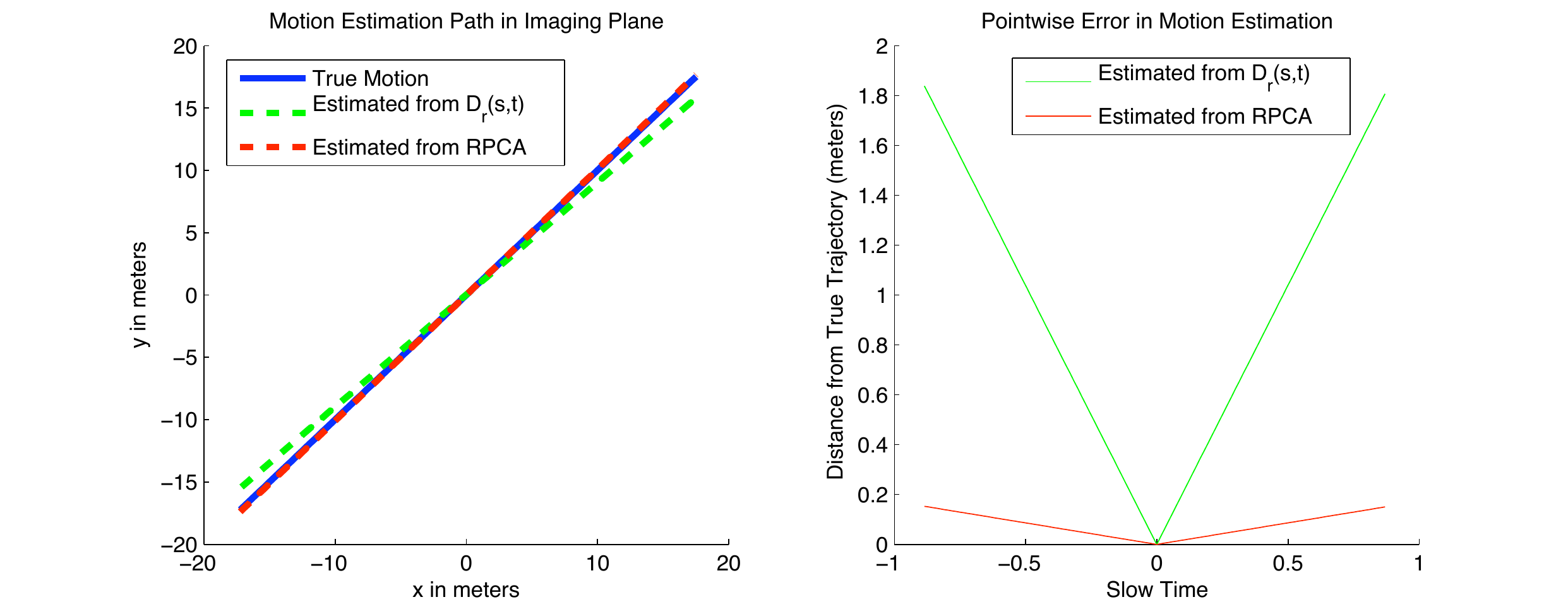}
\caption{Estimation of the trajectory of a moving target in a complex
  scene with 30 stationary scatterers, placed at random in a $50\times
  50\ m^2$ imaging region.  The reflectivity of the moving target is
  ten times stronger than that of the stationary ones.  Left:
  estimated target trajectory using the original traces (green) and
  the separated traces (red). The true trajectory is in blue.  Right:
  errors of the estimated trajectories.}
\label{fig:motEstStrongMov}
\end{figure}

The data separation results are in Figure \ref{fig:rpcaStrongMov}.  We
show in Figure \ref{fig:motEstStrongMov} the estimated target
trajectory with the original data traces (left plot in Figure
\ref{fig:rpcaStrongMov}), and the sparse component (right plot in
Figure \ref{fig:rpcaStrongMov}). We obtain as before that the estimation 
is better after the data separation.

\begin{figure}[!h]
\hspace{-.2in}
\subfigure{\includegraphics[width=.37\columnwidth]{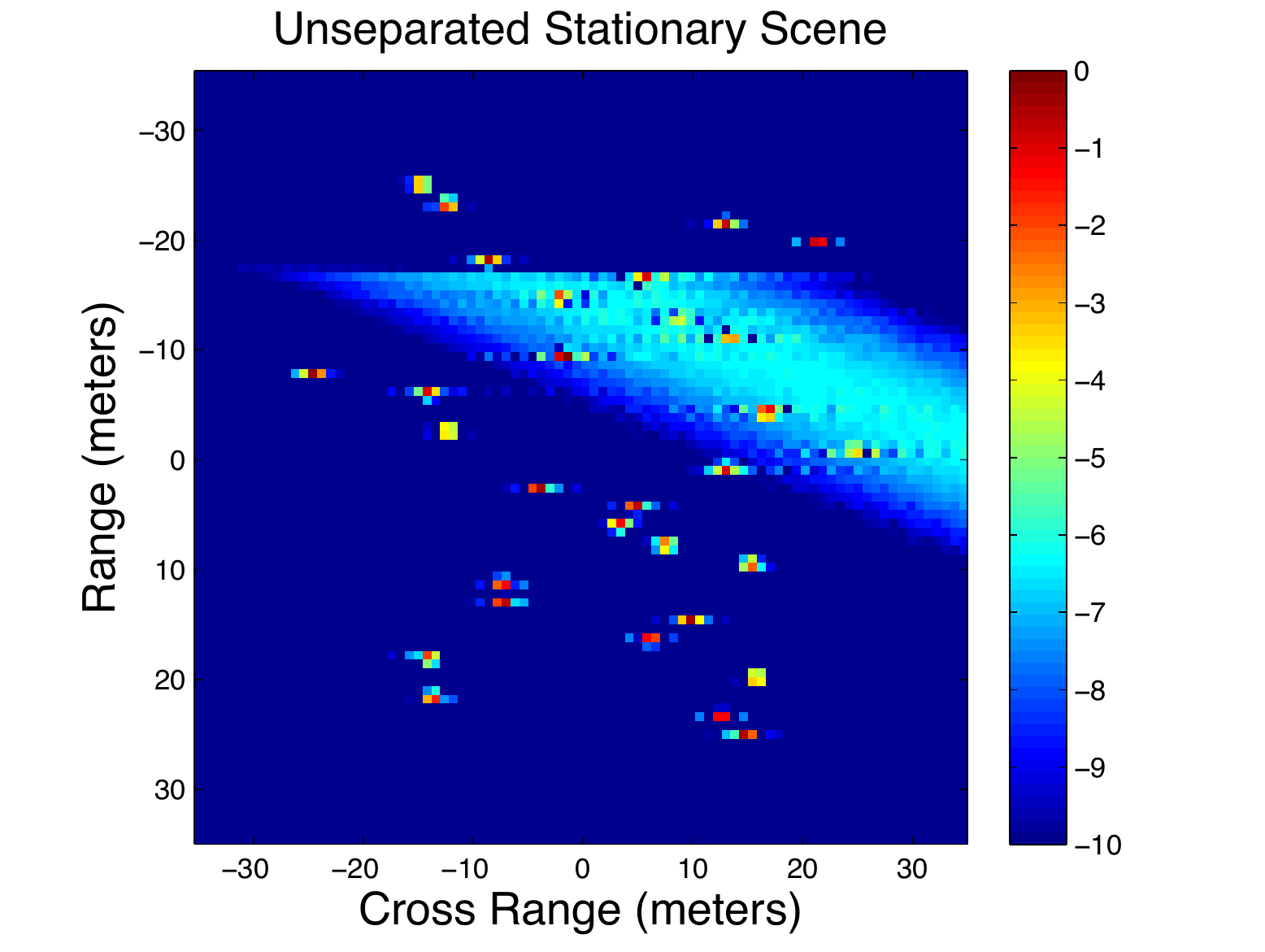}}\hspace{-.25in}
\subfigure{\includegraphics[width=.37\columnwidth]{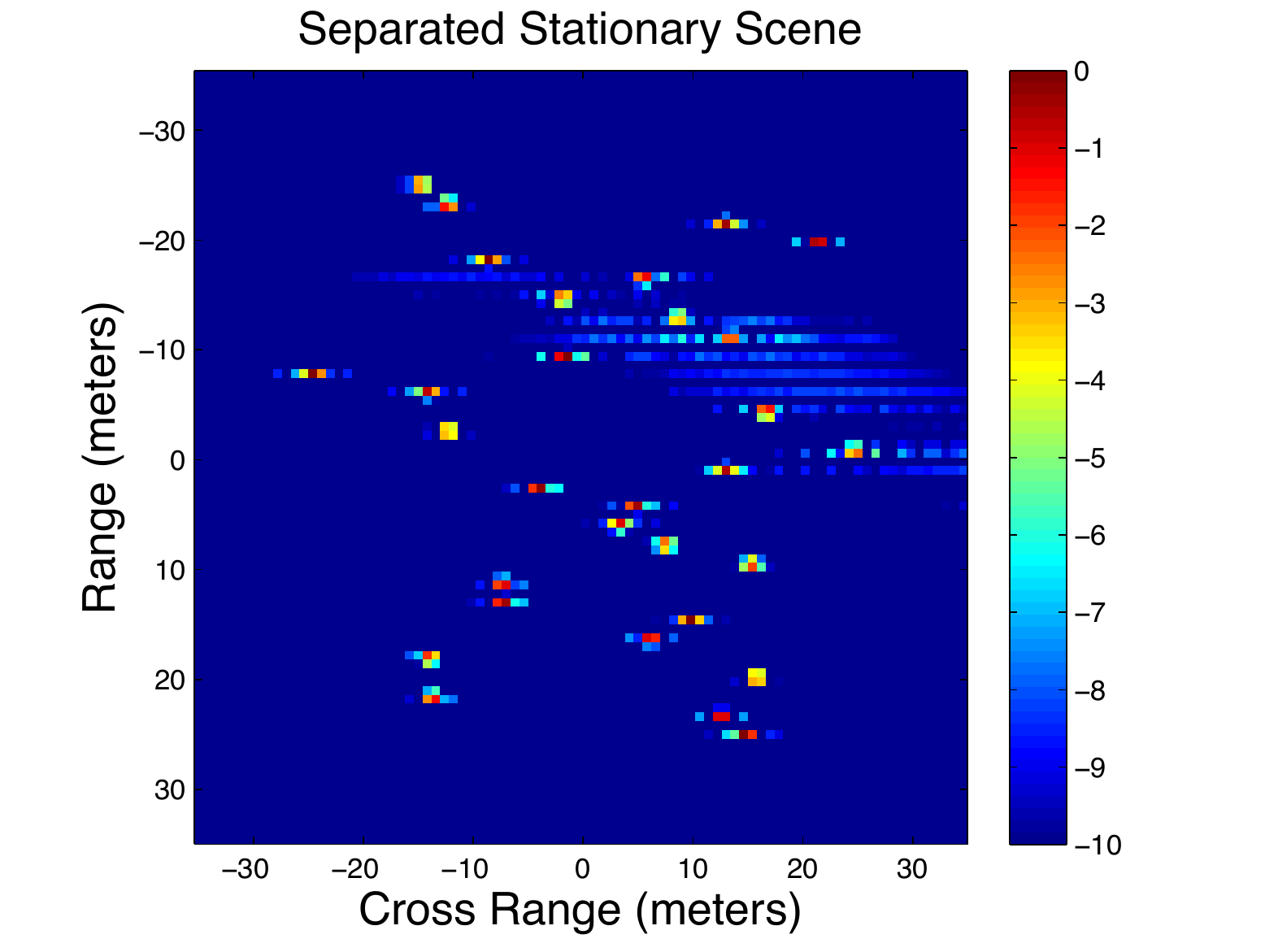}}\hspace{-.25in}
\subfigure{\includegraphics[width=.37\columnwidth]{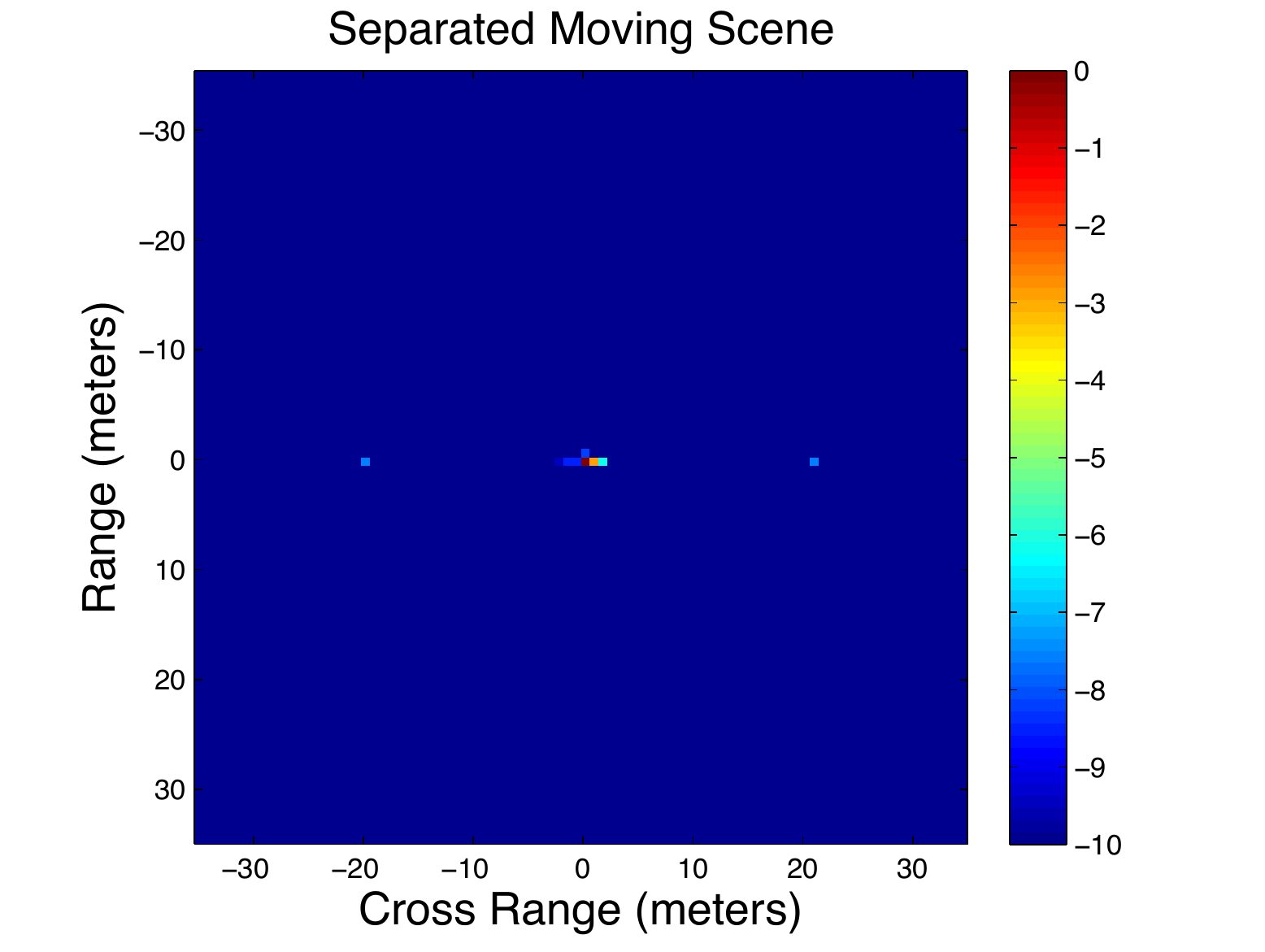}}
\caption{Images of a scene with 30 stationary targets and a moving one
  with reflectivity that is ten times stronger than the others.  Left:
  image obtained with the original data traces.  Middle: image
  obtained with the low rank component of the data traces, returned by
  robust PCA. Right: image obtained with the sparse component of the
  data traces, returned by robust PCA. The motion compensation is with
  the estimated target velocity.}
\label{fig:imagesStrongMov}
\end{figure}

The left plot in Figure \ref{fig:imagesStrongMov} shows the image
computed with the original SAR data traces. The image is focused at
the stationary targets, but there is a strong artifact (streak), due
to the moving target. The middle plot in  Figure \ref{fig:imagesStrongMov}
shows the image computed with the low rank component of the data traces, 
displayed in the middle in Figure \ref{fig:rpcaStrongMov}. The effect 
of the moving target is now considerably smaller. The right plot 
in Figure \ref{fig:imagesStrongMov} shows the image obtained with the 
sparse component of the traces, with motion compensation using the estimated 
target velocity. There is no artifact due to the stationary targets and 
the image is focused at the location $(0,0,0)$m, as expected. 

\section{Summary}
\label{sect:conc}
\setcounter{equation}{0} In this paper we consider the problem of
synthetic aperture radar (SAR) imaging of complex scenes consisting of
a few moving and many stationary targets. The SAR setup is the
usual one with a single antenna mounted on a platform flying above the
region to be imaged. With large bandwidth probing signals and long data
acquisition trajectories, SAR can produce high resolution images of
stationary scenes. However, the presence of moving targets may cause
serious degradation of the images. When the targets have moderate
speed, they appear displaced and blurred in the images.  Fast moving
targets create significant artifacts such as prominent streaks.

To bring the images of the moving targets in focus we need to estimate
their motion. This is necessarily done with small successive
sub-apertures, corresponding to short acquisition times over which the
target motion can be approximated by uniform translation. Imaging with
motion estimation is difficult for at least the following reasons:
First, the echoes from the moving targets may be overwhelmed by those
from the stationary scenes, so the targets may be difficult to
detect. Second, even if we can detect the presence of a moving target
in a complex scene, it is difficult to estimate its motion with the
existing algorithms, unless we use multiple receiver or transmitter
antennas. The algorithms that work with the usual SAR setup assume
that all the targets move the same way and are sensitive to the
presence of strong stationary targets. Third, even if we detect and
estimate well the target motion, when we compensate for it in the
image formation process we may bring the stationary targets out of focus, and thus
still get images with significant artifacts.

To address these challenges, we propose a pre-processing step of the
SAR data designed to separate the stationary target echoes from those
due to the moving targets. The main result of the paper is to show
that this can be accomplished with the robust principal component
analysis (PCA) algorithm complemented with appropriate data windowing.
The robust PCA algorithm decomposes a matrix $M$ into a low rank part
$\cL$ and a sparse part $\cS$. In our context, the matrix $M$ is given
by the pulse and range compressed echoes received at the SAR platform.
We show with analysis and numerical simulations that the contribution
of the stationary targets to $M$ is a low rank matrix when we observe
it in a small enough time window. Thus, we may think of it as the
component $\cL$ of $M$.  The contribution of a few moving targets to
$M$ is a sparse matrix that has higher rank, depending on the target
velocity. Therefore, we expect that robust PCA separates it from the
low rank part $\cL$, due to the stationary targets.

We show with numerical simulations that indeed, robust PCA can
accomplish such data separation. But the algorithm cannot be applied
as a black box.  It must be complemented with proper windowing of the
pulse and range compressed SAR data in order to achieve a good
separation. We present results for various imaging scenes containing
multiple stationary targets, and one or two moving targets that may be
stronger or weaker than the stationary ones. For weaker targets, we
demonstrate that robust PCA can detect their faint echoes and it
separates them from those due to the stationary targets.  We also show
motion estimation and imaging results with and without the data
separation step, in order to demonstrate its importance in achieving
good results.

\section*{Acknowledgement}

The work of L. Borcea was partially supported by the AFSOR Grant
FA9550-12-1-0117, by Air Force-SBIR FA8650-09-M-1523, the ONR Grant
N00014-12-1-0256, and by the NSF Grants DMS-0907746, DMS-0934594.  The
work of T. Callaghan was partially supported by Air Force-SBIR
FA8650-09-M-1523 and the NSF VIGRE grant DMS-0739420.
The work of G. Papanicolaou was supported in part by AFOSR grant FA9550-11-1-0266.

\section*{Appendices}

\def\theequation{{\Alph{section}}.{\arabic{equation}}} 

\appendix

\section{Single target covariance matrix}
\label{sect:appSingle}
\setcounter{equation}{0}

We approximate here the function 
\begin{align*}
  \cC(s,s')&\approx \frac{1}{\Delta t}\int_{-\infty}^{\infty}
  dt\, D_r(s,t)D_r(s',t)\\
  &=\frac{1}{\Delta t}\int_{-\infty}^{\infty} dt\
  \cos(\omega_o(t-\Delta\tau(s,\vrho(s)))) \exp
  \left[-\frac{B^2}{2}(t-\Delta\tau(s,\vrho(s)))^2\right] \\
  & \hspace{0.6in} \times \cos(\omega_o(t-\Delta\tau(s',\vrho(s')))) \exp
  \left[-\frac{B^2}{2}(t-\Delta\tau(s',\vrho(s')))^2\right],
\end{align*}
and simplify notation as
\begin{align*}
  \Delta\tau^-(s,s')&:=\Delta\tau(s,\vrho(s))-\Delta\tau(s',\vrho(s'))\\
  \Delta\tau^+(s,s')&:=\Delta\tau(s,\vrho(s))+\Delta\tau(s',\vrho(s'))
\end{align*}
We rewrite the integrand using a trigonometric identity, and
completing the square
\begin{eqnarray*}
  \cos[\omega_o(t-\Delta\tau(s,\vrho(s)))]
  \cos[\omega_o(t-\Delta\tau(s',\vrho(s)))] \exp\left[-\frac{B^2}{2}
    (t-\Delta\tau(s,\vrho(s)))^2-
    \frac{B^2}{2}(t-\Delta\tau(s,\vrho(s)))^2\right] = \\
  \frac{1}{2}\exp
  \left[-
    \frac{B^2\left(\Delta\tau^-(s,s')\right)^2}{4} \right] \left\{\cos[\omega_0\Delta\tau^-(s,s')]+
    \cos\left[2\omega_o\left(t+\frac{\Delta\tau^+(s,s')}{2}\right)
    \right]\right\}  \\
  \times \exp
  \left[-B^2\left(t+\frac{\Delta\tau^+(s,s')}{2}\right)^2\right],
\end{eqnarray*}
and obtain that 
\begin{eqnarray*}
  \cC(s,s')&=&\frac{1}{2\Delta t}\cos[\omega_0\Delta\tau^-(s,s')]
  \exp \left[- \frac{B^2\left(\Delta\tau^-(s,s')\right)^2}{4} \right]
  \int_{-\infty}^{\infty} dt\
  \exp \left[-B^2\left(t+\frac{\Delta\tau^+(s,
        s')}{2}\right)^2 \right]\\
  && +
  \frac{1}{2\Delta t}\exp\left[-
    \frac{B^2\left(\Delta\tau^-(s,s')\right)^2}{4}\right]
  \int_{-\infty}^{\infty} dt\ \cos\left[2\omega_o\left(t+
      \frac{\Delta\tau^+(s,s')}{2}\right)\right]\exp \left[-B^2
    \left(t+\frac{\Delta\tau^+(s,s')}{2}\right)^2\right]\\
  &=&\frac{\sqrt{\pi}}{2B\Delta t}\cos[
  \omega_0\Delta\tau^-(s,s')]\exp \left[-\frac{B^2\left(
        \Delta\tau^-(s,s')\right)^2}{4}\right]-
  \frac{\sqrt{\pi}}{2B\Delta t}\exp\left[-\frac{B^2
      \left(\Delta\tau^-(s,s')\right)^2}{4}-
    \frac{\omega_o^2}{B^2}\right]\\
  &\approx&\frac{\sqrt{\pi}}{2B\Delta
    t}\cos[\omega_0\Delta\tau^-(s,s')]\exp \left[
-\frac{B^2\left(\Delta\tau^-(s,s')\right)^2}{4}\right].
\end{eqnarray*}
The last approximation is because in our regime $\om_o \gg B$.  

Our assumption (\ref{eq:Fresnel}) on the Fresnel number, and therefore
on the aperture, allows us to linearize $\Delta \tau^{-}(s,s')$ and
obtain the Toeplitz structure stated in Proposition \ref{prop:Toep}.
We have by the mean value theorem that
\begin{equation}
  \Delta\tau^-(s,s') = (s-s') \frac{d}{ds} \Delta \tau(\bar{s},\vrho(
\bar{s}))
\end{equation}
for some $\bar s$ between $s$ and $s'$. The derivative is given by 
\begin{equation}
  \frac{d}{ds} \Delta \tau(\bar{s},\vrho(
  \bar{s})) = \frac{2}{c_o} \left[V \vt(\bar s) \cdot \left(\vm(\bar s)
      -\vm_o(\bar s)\right) 
    - \vu \cdot \vm(\bar s) \right],
\label{eq:DD}
\end{equation}
in terms of the unit vectors 
\[
\vm(\bar s) = \frac{\vr(\bar s) - \vrho(\bar s)}{|\vr(\bar s) -
  \vrho(\bar s)|}, \quad 
\vm_o(\bar s) = \frac{\vr(\bar s) - \vrho_o}{|\vr(\bar s) -
  \vrho_o|},
\]
and the unit vector $\vt(\bar s)$ tangential to the flight path at
$\vr(\bar s)$. We use that 
\[
\vm(\bar s) -\vm_o(\bar s) = \left[I-\vm_o(\bar s)\vm_o(\bar
  s)^T\right] \frac{(\vrho(\bar s) - \vrho_o)}{|\vr(\bar s) -
  \vrho_o|} + O \left[ \left( \frac{R^{^\cI}}{L} \right)^2 \right],
\]
and expand the right hand side in (\ref{eq:DD}) around $\bar s = 0$
and obtain
\begin{equation}
  \frac{d}{ds} \Delta \tau(\bar{s},\vrho(
  \bar{s})) = \frac{2}{c_o} \left[V \vt \cdot \frac{\Pp_o 
      (\vrho - \vrho_o)}{|\vr(0) -
      \vrho_o|} - \vu \cdot \vm_o - \vu \cdot \frac{\Pp_o 
      (\vrho - \vrho_o)}{|\vr(0) - \vrho_o|}
  \right] + O\left(\frac{a \vt \cdot \Pp_o \vu}{c_oL}\right) + O
  \left(\frac{V a R^{^\cI}}{c_oL^2}\right).
\end{equation}
Therefore, 
\begin{equation}
  \om_o \Delta\tau^-(s,s') = 
  \frac{2(s-s')}{c_o} \left[V \vt \cdot \frac{\Pp_o (\vrho - 
      \vrho_o)}{|\vr(0) -
      \vrho_o|} - \vu \cdot \vm_o - \vu \cdot \frac{\Pp_o (\vrho 
      - \vrho_o)}{|\vr(0) -
      \vrho_o|}
  \right] + \mathcal{E},
\end{equation}
with negligible error by assumption (\ref{eq:Fresnel})
\[
\mathcal{E} = O\left(\frac{a^2 \vt \cdot \Pp_o \vu}{\la_o L V}\right)
+ O \left(\frac{a^2 R^{^\cI}}{\la_o L^2}\right) \ll 1.
\]
This is the result stated in Proposition \ref{prop:Toep}.

\section{Computation of the symbol}
\label{sect:appSymbol}
\setcounter{equation}{0}
The symbol is given by
\begin{equation}
  Q(\theta) = \sum_{j=-\infty}^{\infty} c_j e^{ij\theta},\quad\quad \theta\in (-\pi,\pi)
\end{equation}
with $c_j$ defined by (\ref{eq:defcj}).  Thus
\begin{align*}
Q(\theta)&= \frac{\sqrt{\pi}}{2B\Delta t}\sum_{j=-\infty}^{\infty}
e^{-\frac{(\xi j)^2}{4}}\cos(\gamma j) e^{ij\theta}\\ &=
\frac{\sqrt{\pi}}{2B\Delta t}\left(\sum_{j=-\infty}^{\infty} \left[
  \cos(j\theta)\cos(\gamma j)e^{-\frac{(\xi
      j)^2}{4}}\right]+i\sum_{j=-\infty}^{\infty}
\left[\sin(n\theta)\cos(\gamma j)e^{-\frac{(\xi
      j)^2}{4}}\right]\right)\\ &= \frac{\sqrt{\pi}}{4B\Delta
  t}\sum_{j=-\infty}^{\infty} \left[
  \cos(n(\theta-\gamma))+\cos(n(\theta+\gamma))\right]e^{-\frac{(\xi
    j)^2}{4}},
\end{align*}
where we recall that 
\[
\xi = B |\alpha| \Delta s.
\]
Define $\Delta x=\xi/2$ and $x_j=j\Delta x$.  Then
\begin{align*}
Q(\theta)&=\frac{\sqrt{\pi}}{4B\Delta t}\frac{1}{\Delta x}
\sum_{j=-\infty}^{\infty} \left[ \cos\left(\frac{2(\theta-\gamma
    )}{\xi}x_j\right)+ \cos\left(\frac{2(\theta+\gamma
    )}{\xi}x_j\right)\right]e^{-x_j^2}\Delta x
\end{align*}
For $\theta\in(-\pi,\pi)$ and $\xi$ (i.e., $\Delta s$) small enough,
we can approximate the sum with an integral
\begin{align*}
Q(\theta)&\approx \frac{\sqrt{\pi}}{2B\Delta
  t\xi}\int_{-\infty}^{\infty} \left[ \cos\left(\frac{2(\theta-\gamma
    )}{\xi}x\right)+\cos\left(\frac{2(\theta+\gamma
    )}{\xi}x\right)\right]e^{-x^2}dx\\ &= \frac{\pi}{2B\Delta
  t\xi}\left[e^{-\frac{(\theta-\gamma)^2}{\xi^2}}+
  e^{-\frac{(\theta+\gamma)^2}{\xi^2}}\right].
\end{align*}

\section{Rank estimate of large Toeplitz plus g-Hankel matrices}
\label{sect:gHankel}
\setcounter{equation}{0}

We use the results from \cite{sesana} to obtain the asymptotic
estimate of the rank of matrix $C$ given by equation (\ref{eq:sumTH})
as the sum of a Toeplitz matrix $T$, a g-Hankel matrix $H$ and its
transpose.  We need the following definition:
\begin{definition}
\label{def.2}
Let $Q$ be a complex valued, measurable function defined on the
interval $U = (-\pi,\pi)$. Let also $\{A_n\}_{n \in \mathbb{N}}$ be a
sequence of of matrices. Each matrix $A_n \in \mathbb{R}^{(n+1) \times
  (n+1)}$, and we denote by $\sigma_j(A_n)$ its singular values, in
descending order, for $j = 1, \ldots, n+1$.  We say that the sequence
is distributed (in the sense of singular values) as the pair $(Q,U)$,
and write in short $\{A_n\} \sim_{\sigma}(Q,U)$, if
\begin{equation}
\lim_{n\to\infty} \frac{1}{n+1} \sum_{j=1}^{n+1}
F(\sigma_j(A_n))=\frac{1}{2\pi}\int_{-\pi}^{\pi}F(|Q(\theta)|)d\theta,
\end{equation}
for every $F \in C_o(\mathbb{R}^+)$. Here $C_o(\mathbb{R}^+)$ is the
set of continuous functions with bounded support over the nonnegative
real numbers.
\end{definition}

It is shown in \cite[section 4.2.2]{sesana} that if $\{H_n\}_{n \in \mathbb{N}}$ 
is a sequence of g-Hankel matrices $H_n \in \mathbb{R}^{(n+1) \times
  (n+1)}$, then 
\begin{equation}
\{H_n\}\sim_{\sigma}(0,U).
\label{eq:RES.1}
\end{equation}
Moreover, \cite[Proposition 4.3]{sesana} states that if $\{A_n\}_{n
  \in \mathbb{N}}$ and $\{H_n\}_{n\in \mathbb{N}}$ are two sequences
of matrices satisfying 
$\{A_n\}\sim_{\sigma}(Q,U)$ and $\{H_n\}\sim_{\sigma}(0,U)$ then
\begin{equation}
\{A_n+H_n\}\sim_{\sigma}(Q,U).
\label{eq:RES.2}
\end{equation}
In our context, $A_n = T_n$ are Toeplitz matrices defined by the
sequence $\{c_j\}_{j \in \mathbb{Z}}$, and $Q$ is the symbol defined by
(\ref{eq:series}). We are interested in the distribution (in the sense
of singular values) of the sequence $\{C_n\}_{n \in \mathbb{N}}$ defined
by
\begin{equation}
C_n = T_n + H_n + H_n^T.
\end{equation}
Because the singular values of the transpose $H_n^T$ are the same as
the singular values of $H_n$, we obtain using (\ref{eq:RES.1}) and 
Definition (\ref{def.2}) that 
\begin{equation}
\{H_n^T\}\sim_{\sigma}(0,U).
\label{eq:RES.3}
\end{equation}
Thus, $\{(T_n + H_n) + H_n^T\}$ has the same distribution as $\{T_n +
H_n\}$ and by (\ref{eq:RES.2}),
\begin{equation}
\{T_n + H_n\} \sim_\sigma (Q,U).
\label{eq:RES.4}
\end{equation}
More explicitly, 
\begin{equation}
\lim_{n\to\infty} \frac{1}{n+1} \sum_{j=1}^{n+1}
F(\sigma_j(T_n+H_n+H_n^T))=\frac{1}{2\pi}\int_{-\pi}^{\pi}F(|Q(\theta)|)d\theta,
\qquad \forall F \in C_0(\mathbb{R}^+).
\label{eq:RES.5}
\end{equation}

We cannot apply directly this result to the computation of rank,
because the indicator function $1_{[\epsilon \|Q\|_{\infty}, \infty)}$
  does not have bounded support and it is not continuous. However, the
  result extends to such functions as we shown next.

First, let us show that the sigular values of matrices $T_n + H_n$ are
bounded uniformly in $n$. Because the largest singular value of a
matrix is equal to its 2-norm, we obtain by the triangle inequality
that
\[
\sigma_1(T_n + H_n + H_n^T) = \|T_n + H_n + H_n^T\|_2 \le \|T_n \|_2 + 2 \|H_n\|_2 =
\sigma_1(T_n) + 2\sigma_1(H_n).
\] 
One of the results of the Szeg\H{o} theory for large Toeplitz
matrices \cite{szego,bottcher} is that the sequence of largest
singular values $\{\sigma_1(T_n)\}_{n \in \mathbb{N}}$ converges to
the limit $\|Q\|_{\infty}$. Thus, the sequence is bounded above, and we
denote the bound by $\Sigma_T$.  For the g-Hankel matrix we can use 
the matrix norm inequality 
\[
\sigma_1(H_n) = \|H_n\|_2 \le \sqrt{ \|H_n\|_1 \|H_n\|_\infty},
\]
where $\|H_n\|_1$ and $\|H_n\|_\infty$ are equal to the maximum of the
1-norm of the columns and rows of $H_n$, respectively. They are
defined by equation (\ref{eq:Hg}), in terms of the sequence $\{h_j\}$
given in equation (\ref{eq:seqh}). Obviously, the 1-norm of the columns 
and rows of $H_n$ are bounded above by the series 
\[
\Sigma_H = \sum_{j=0}^\infty |h_j| < \infty,
\]
which is convergent because $h_j$ decays exponentially with $j$.
Therefore, 
\[
\|H_n \|_1 \le \Sigma_H \quad \mbox{and} \quad  \|H_n \|_\infty \le \Sigma_H
\]
and gathering the results above, we have that
\begin{equation}
\sigma_1(T_n + H_n) \le \Sigma_T + 2 \Sigma_H.
\label{eq:BOUND}
\end{equation}

Since the singular values are bounded by (\ref{eq:BOUND}), in our
calculation of rank we can replace the indicator function
$1_{[\epsilon \|Q\|_{\infty}, \infty)}(x)$ with a new function
  $\chi(x)$ of bounded support, satisfying
\begin{equation}
\chi(x) = \begin{cases}
0 & ~ ~ x < \delta,\\
1 & ~ ~x \in [\delta, D],
\end{cases}
\end{equation}
and decaying to zero in a continuum manner for $x > D$.  Here we
simplified notation as
\[
\delta:= \epsilon \|Q\|_{\infty} \quad \mbox{and} \quad D = \Sigma_T + 2 \Sigma_H.
\]
It remains to show that result (\ref{eq:RES.5}) extends to the
function $\chi$, which has bounded support but is discontinuous at $x
= \delta$.

Let us introduce the sequence of continuous functions $\{\chi_m(x)\}_{m \in
  \mathbb{Z}^{+}}$, defined by
\begin{equation}
\chi_m(x)=\begin{cases}
0 & ~ ~x<\delta-\frac{1}{m}\\
m(x-\delta)+1 & \delta-\frac{1}{m}\le x\le\delta\\
\chi(x) & ~ ~ x > \delta.
\end{cases}
\end{equation}
This sequence converges pointwise to $\chi(x)$, as $m \to \infty$.  We
know that (\ref{eq:RES.5}) holds for $F = \chi_m$. To extend the
result to $F = \chi$, we show next that we can interchange the limits
as in
\begin{align*}
\lim_{m\to\infty}\lim_{n\to\infty} \frac{1}{n+1}\sum_{j=1}^{n+1}
\chi_m\left(\sigma_j(T_n+H_n +
H_n^T)\right)&=\lim_{n\to\infty}\lim_{m\to\infty}\frac{1}{n+1}\sum_{j=1}^{n+1}
\chi_m \left(\sigma_j(T_n+H_n +
H_n^T)\right)\\ &=\lim_{n\to\infty}\frac{1}{n+1}\sum_{j=1}^{n+1}
\chi\left(\sigma_j(T_n+H_n + H_n^T)\right).
\end{align*}
We have that 
\begin{align}
&\lim_{m\to\infty}\lim_{n\to\infty} \frac{1}{n+1}\sum_{j=1}^{n+1}
  \chi_m\left(\sigma_j(T_n + H_n +
  H_n^T)\right)-\lim_{n\to\infty}\frac{1}{n+1}\sum_{j=1}^{n+1}
  \chi\left(\sigma_j(T_n + H_n +
  H_n^T)\right)\nonumber\\ &=\lim_{m\to\infty}\left[\lim_{n\to\infty}
    \frac{1}{n+1}\sum_{j=1}^{n+1} \left[\chi_m\left(\sigma_j(T_n + H_n
      + H_n^T)\right)-\chi\left(\sigma_j(T_n + H_n +
      H_n^T)\right)\right]\right]\nonumber
  \\ &=\lim_{m\to\infty}\left[\lim_{n\to\infty}
    \frac{1}{n+1}\sum_{j=1}^{n+1} \left[g_m\left(\sigma_j(T_n + H_n +
      H_n^T)\right)\right]\right]
\label{eq:RES}
\end{align}
with residual $g_m:=\chi_m-\chi$ that satisfies by construction
\[
g_m(x) \ge 0, \qquad \forall x \in \mathbb{R},
\]
and is bounded above by the continous function
\[
{\mathcal G}_m(x)=\begin{cases}
0 & x<\delta-\frac{1}{m}\\
m(x-\delta)+1 & \delta-\frac{1}{m}\le x\le\delta\\
-m(x-\delta)+1 & \delta< x\le\delta+\frac{1}{m}\\
0 & x>\delta+\frac{1}{m}.
\end{cases}
\]
This gives
\begin{align*}
0 \le \lim_{m\to\infty}\left[\lim_{n\to\infty} \frac{1}{n+1}\sum_{j=1}^{n+1}
  \left[g_m\left(\sigma_j(T_n + H_n + H_n^T)\right)\right]\right]&\le
\lim_{m\to\infty}\left[\lim_{n\to\infty} \frac{1}{n+1}\sum_{j=1}^{n+1}
  \left[{\mathcal G}_m\left(\sigma_j(T_n + H_n + H_n^T)\right)\right]\right]\\ &=
\lim_{m\to\infty}\frac{1}{2\pi}\int_{-\pi}^{\pi}
{\mathcal G}_m\left(Q(\theta)\right)d\theta,
\end{align*}
and we can bring the limit inside the integral using the dominated
convergence theorem
\[
\lim_{m\to\infty}\frac{1}{2\pi}\int_{-\pi}^{\pi}
{\mathcal G}_m\left(Q(\theta)\right)d\theta = 
\frac{1}{2\pi}\int_{-\pi}^{\pi}
\lim_{m\to\infty} {\mathcal G}_m\left(Q(\theta)\right)d\theta =0.
\]
The last equality is because ${\mathcal G}_m \to 0$ almost everywhere.
Thus, the limit (\ref{eq:RES}) is equal to zero, and the result
follows.

\bibliographystyle{plain} \bibliography{SBIR}

\end{document}